\documentclass[structabstract]{aa}  
\usepackage{graphicx}
\usepackage{amsmath}
\usepackage{amsfonts}
\usepackage{amssymb}
\usepackage{algorithm,algorithmic}
\usepackage{txfonts}
\usepackage{natbib}
\bibpunct{(}{)}{;}{a}{}{,}

\begin{document}

\title{A conjugate gradient algorithm \\ for the astrometric core solution of Gaia}
\titlerunning{A conjugate gradient algorithm for Gaia}

\author{
A. Bombrun\inst{1}
\and L. Lindegren\inst{2}
\and D. Hobbs\inst{2}
\and B. Holl\inst{2}
\and U. Lammers\inst{3}
\and U. Bastian\inst{1}
}

\institute{Astronomisches Rechen-Institut, Zentrum f{\"u}r Astromomie der Universit{\"a}t
Heidelberg, M{\"o}nchhofstr. 12--14, DE-69120 Heidelberg, Germany \\
\email{abombrun@ari.uni-heidelberg.de, bastian@ari.uni-heidelberg.de}
\and
Lund Observatory, Lund University, Box 43, SE-22100 Lund, Sweden\\
\email{lennart@astro.lu.se, berry@astro.lu.se, david@astro.lu.se}
\and
European Space Agency (ESA), European Space Astronomy Centre (ESAC),
P.O. Box (Apdo.~de~Correos) 78, ES-28691 Villanueva de la Ca{\~n}ada, Madrid, Spain\\
\email{Uwe.Lammers@sciops.esa.int}
}

\date{Received 17 August 2011 / Accepted 25 November 2011}

\abstract{
The ESA space astrometry mission Gaia, planned to be launched in 2013, 
has been designed
to make angular measurements on a global scale with micro-arcsecond accuracy. A key
component of the data processing for Gaia is the astrometric core solution, which must
implement an efficient and accurate numerical algorithm to solve the resulting, extremely
large least-squares problem. The Astrometric Global Iterative Solution (AGIS) is a
framework that allows to implement a range of different iterative solution schemes 
suitable for a scanning astrometric satellite.
}
{
Our aim is to find a computationally efficient and numerically accurate iteration scheme
for the astrometric solution, compatible
with the AGIS framework, and a convergence criterion for deciding when to stop 
the iterations.
}
{
We study an adaptation of the classical conjugate gradient (CG) algorithm, and 
compare it to
the so-called simple iteration (SI) scheme that was previously known to 
converge for this problem, although
very slowly. The different schemes are implemented within a software test bed for AGIS
known as AGISLab. This allows to define, simulate and study scaled astrometric core
solutions with a much smaller number of unknowns than in AGIS, and therefore to perform a
large number of numerical experiments in a reasonable time.
After successful testing in AGISLab, the CG scheme has been implemented also in AGIS.
}
{
The two algorithms CG and SI eventually converge to identical solutions, to
within the numerical noise (of the order of 0.00001~micro-arcsec). These 
solutions are moreover independent of the starting values (initial star catalogue),
and we conclude that they are equivalent to a rigorous least-squares estimation 
of the astrometric parameters. The CG scheme converges up to a factor four faster 
than SI in the tested cases, and in particular spatially correlated truncation errors 
are much more efficiently damped out with the CG scheme. While it appears to be
difficult to define a strict and robust convergence criterion, we have found that the 
sizes of the updates, and possibly the correlations between the updates in successive 
iterations, provide useful clues. 
}
{}

\keywords{
Astrometry --
Methods: data analysis --
Methods: numerical --
Space vehicles: instruments
}

\maketitle

\section{Introduction}
\label{sec-intro}

The European Space Agency's Gaia mission \citep{gaia2001,lind+08,Lindegren10} 
is designed to measure
the astrometric parameters (positions, proper motions and parallaxes) of around one
billion objects, mainly stars belonging to the Milky Way Galaxy and the local group.
The scientific processing of the Gaia observations is a complex task that requires
the collaboration of many scientists and engineers with a broad range of expertise
from software development to CCDs. A consortium of European research centres and
universities, the Gaia Data Processing and Analysis Consortium (DPAC), has been set
up in 2005 with the goal to design, implement and operate this process 
\citep{mign+08}. In this paper we focus on a central component of the scheme, 
namely the astrometric core solution, which solves the
corresponding least-squares problem within a software framework known as the
Astrometric Global Iterative Solution, or AGIS \citep{Lammers+09,agis11,WOM2010}.

In a single solution, the AGIS software will simultaneously calibrate the instrument,
determine the three-dimensional orientation (attitude) of the instrument as a function 
of time, produce the catalogue of astrometric parameters of the stars, and link it to 
an adopted celestial reference frame. This computation is based on the results of
a preceding treatment of the raw satellite data, basically giving the measured
transit times of the stars in the instrument focal plane \citep{Lindegren10}. 
The astrometric core solution can be considered as a least-squares problem with 
negligible non-linearities except for the outlier treatment. Indeed, it should only 
take into account so-called primary sources, that is stars and
other point-like objects (such as quasars) that can astrometrically 
be treated as single stars to the required
accuracy. The selection of the primary sources is a key component of the
astrometric solution, since the more that are used the better the instrument can
be calibrated, the more accurate the attitude can be determined, and the better
the final catalogue will be. This selection, and the identification of outliers
among the individual observations, will be made recursively after reviewing the
residuals of previous solutions \citep{agis11}. 
What remains is then, ideally, a `clean' set of data referring to the
observations of primary sources, from which the astrometric core solution 
will be computed by means of AGIS.

From current estimates, based on the known instrument capabilities and star
counts from a Galaxy model, it is expected that at least 100~million primary
sources will be used in AGIS. Nonetheless, the solution would be strengthened if even
more primary sources could be used. Moreover, it should be remembered that AGIS
will be run many times as part of a cyclic data reduction scheme, where the
(provisional) output of AGIS is used to improve the raw data treatment
\citep[the Intermediate Data Update; see][]{OMull+09}.
Hence, it is important to ensure that AGIS can be run both very efficiently from
a computational viewpoint, and that the end results are numerically accurate,
i.e., very close to the true solution of the given least-squares problem.

Based on the generic principle of self-calibration, the attitude and calibration 
parameters are derived from the same set of observational data as the astrometric
parameters. The resulting strong coupling between the different kinds of
parameters makes a direct solution of the resulting equations extremely 
difficult, or even unfeasible by several orders of magnitude with current 
computing resources \citep{bom-lin-hol-09}. On the other hand, this coupling 
is well suited for a block-wise organization of the equations, where, for example,
all the equations for a given source are grouped together and solved, 
assuming that the relevant attitude and calibration parameters are already
known. The problem then is of course that, in order to compute the astrometric 
parameters of the sources to a given accuracy, one needs to know first the 
attitude and calibration parameters to corresponding accuracies; these in turn
can only be computed once the source parameters have been obtained 
to sufficient accuracy; and so on. This organization of the computations 
therefore naturally leads to an iterative solution process. Indeed, in AGIS 
the astrometric solution is broken down into (at least) three distinct blocks, 
corresponding to the source, attitude and calibration parameter updates, 
and the software is designed to optimize data throughput within this 
general processing framework \citep{Lammers+09}. Cyclically computing 
and applying the updates in these blocks corresponds to the so-called 
simple iteration (SI) scheme (Sect.~\ref{sec-simple}), which is known to
converge, although very slowly. 

However, it is possible to implement many other iterative algorithms 
within this same processing framework, and some of them may 
exhibit better convergence properties than the SI scheme. 
For example, it is possible to speed up the convergence if the updates 
indicated by the simple iterations are extrapolated by a certain factor. 
More sophisticated algorithms could be derived from various 
iterative solution methods described in the literature.

The purpose of this paper is to describe one
specific such algorithm, namely the conjugate gradient (CG) algorithm with 
a Gauss--Seidel preconditioner, and to show how it can be implemented 
within the AGIS processing framework. We want to make it plausible that it 
indeed provides a rigorous solution to the given least-squares problem. 
Also, we will study its convergence properties in comparison to the SI 
scheme and, if possible, derive a convergence criterion for stopping
the iterations.

Our focus is on the high-level adaptation of the CG algorithm to the present
problem, i.e., how the results from the different updating blocks in AGIS 
can be combined to provide the desired speed-up of the convergence. To test this, 
and to verify that the algorithm provides the correct results, we need to
conduct many numerical experiments, including the simulation of input 
data with well-defined statistical properties, and iterate the solutions 
to the full precision allowed by the computer arithmetic. On the other hand, 
since it is not our purpose to validate the detailed source, instrument and 
attitude models employed by the updating blocks, we can accept a number 
of simplifications in the modelling of the data, such that the experiments 
can be completed in a reasonable time. 
The main simplifications used in the present study are as follows:
\begin{enumerate}
\item
For conciseness we limit the present study to the source and attitude 
parameters, whose mutual disentanglement is by far the most critical 
for a successful astrometric solution \citep[cf.][]{bom-lin-hol-09}. 
For the final data reduction many calibration parameters must also be
included, as well as global parameters
\citep[such as the PPN parameter $\gamma$;][]{Hobbs+09}, and possibly 
correction terms to the barycentric velocity of Gaia derived from stellar 
aberration \citep{Butkevich+08}. These extensions, within the CG scheme, 
have been implemented in AGIS but are not considered here.
\item
We use a scaled-down version of AGIS, known as AGISLab
(Sect.~\ref{sec-agislab}), which
makes it possible to generate input data and perform solutions with a 
much smaller number of primary sources than would be required for the 
(full-scale) AGIS system. This reduces computing time by a large factor, 
while retaining the strong mutual entanglement of the source and attitude
parameters, which is the main reason why the astrometric solution is so
difficult to compute.
\item
The rotation of the satellite is assumed to follow the so-called nominal
scanning law, which is an analytical prescription for the pointing of the 
Gaia telescopes as a function of time. That is, we ignore the small 
($< 1$~arcmin) pointing errors that the real mission will have, 
as well as attitude irregularities, data gaps, etc. The advantage is that 
the attitude modelling becomes comparatively simple and can use a smaller 
set of attitude parameters, compatible with the scaled-down 
version of the solution.
\item
The input data are `clean' in the sense that there are no outliers, and
the observation noise is unbiased with known standard deviation. This
highly idealised condition is important in order to test that the solution 
itself does not introduce unwanted biases and other distortions of the results.
\end{enumerate}
An iterative scheme should in each iteration compute a better approximation
to the exact solution of the least-squares problem. In this paper 
we aim to demonstrate 
that the SI and CG schemes are converging in the sense that the errors,
relative to an exact solution, vanish for a sufficient number of iterations. 
Since we work with simulated data, 
we have a reference point in the true values of the source parameters (positions, 
proper motions and parallaxes) used 
to generate the observations. We also aim to demonstrate that the CG method 
is an efficient scheme to solve the astrometric least-squares problem,
i.e., that it leads, in a reasonable number of iterations, to an approximation 
that is sufficiently close to the exact solution. An important problem when using
iterative solution methods is how to know when to stop, and we study some
possible convergence criteria with the aim to reach the maximum possible 
numerical accuracy.

The paper provides both a detailed presentation of the SI and CG algorithms 
at work in
AGIS and a study of their numerical behaviour through the use of the AGISLab
software \citep{Holl+09}. The paper is organized as follows: Section~\ref{sec-iter}
gives a brief overview of iterative methods to solve a linear least-squares
problem. Section~\ref{sec-algo} describes in detail the algorithms considered
here, viz., the SI and CG with different preconditioners. In Sect.~\ref{sec-conv}
we analyze the convergence of these algorithms and some properties of the
solution itself. Then, Sect.~\ref{sec-agis} presents the
implementation status of the CG scheme in AGIS before the main
findings of the paper are summarized in the concluding
Sect.~\ref{sec-conc}.

\section{Iterative solution methods}
\label{sec-iter}

This section presents the mathematical basis of the simple iteration and 
conjugate gradient algorithms to solve the linear least-squares problem.
For a more detailed description of these and other iterative solution methods 
we refer to \citet{book:bjork-1996} and \citet{vanVorst:03}. A history of 
the conjugate gradient method can be found in \citet{golub:89}.

Let $\vec{M}\vec{x}=\vec{h}$ be the overdetermined set of observation (design) 
equations, where $\vec{x}$ is the vector of unknowns, $\vec{M}$ the design 
matrix, and $\vec{h}$ the right-hand side of the design equations. The unknowns 
are assumed to be (small) corrections to a fixed set of reference values for 
the source and attitude parameters. These reference values must be close 
enough to the exact solution that non-linearities in $\vec{x}$ can be neglected; 
thus $\vec{x}=\vec{0}$ is still within the linear regime. Moreover, we assume 
that the design equations have been multiplied by the square root of their 
respective weights, so that they can be treated by ordinary (unweighted) 
least-squares. That is, we seek the vector $\vec{x}$ that minimizes the sum
of the squares of the design equation residuals,
\begin{equation}
Q = \|\vec{h}-\vec{M}\vec{x}\|^2 \, ,
\end{equation}
where $\|\!\cdot\!\|$ is the Euclidean norm. 
It is well known (cf.\ Appendix~\ref{sec-gauss-markoff}) that if $\vec{M}$ has full 
rank, i.e., $\|\vec{M}\vec{x}\|>0$ for all $\vec{x}\neq\vec{0}$, this problem has a 
unique solution that can be obtained by solving the normal equations 
\begin{equation}\label{eq:ne}
\vec{N}\vec{x} = \vec{b} \, , 
\end{equation}
where $\vec{N}=\vec{M}'\vec{M}$ is the normal matrix,
$\vec{M}'$ is the transpose of $\vec{M}$, and $\vec{b}=\vec{M}'\vec{h}$ the
right-hand side of the normals.
This solution is denoted $\hat{\vec{x}}=\vec{N}^{-1}\vec{b}$. 
In the following, the number of unknowns is denoted $n$ and the number of 
observations $m\gg n$. Thus $\vec{M}$, $\vec{x}$ and $\vec{h}$ have dimensions 
$m\times n$, $n$ and $m$, respectively, and $\vec{N}$ and $\vec{b}$ have 
dimensions $n\times n$ and $n$.

The aim of the iterative solution is to generate a sequence of approximate
solutions $\vec{x}_0$, $\vec{x}_1$, $\vec{x}_2$, $\dots$, such that
$\|\vec{\epsilon}_k\|\rightarrow 0$ as $k\rightarrow\infty$,
where $\vec{\epsilon}_k=\vec{x}_k-\hat{\vec{x}}$ is the truncation error in iteration $k$. 
The design equation residual vector at this point is denoted 
$\vec{s}_k=\vec{h}-\vec{M}\vec{x}_k$ (of dimension $m$), and the normal equation 
residual vector is denoted $\vec{r}_k=\vec{b}-\vec{N}\vec{x}_k=-\vec{N}\vec{\epsilon}_k$ 
(of dimension $n$). 
The least-squares solution $\hat{\vec{x}}$ corresponds to $\hat{\vec{r}}=\vec{0}$. 
At this point we still have in general $\|\hat{\vec{s}}\|>0$, since the design 
equations are overdetermined. If $\vec{x}^\text{(true)}$ are the true 
parameter values, we denote by $\vec{e}_k=\vec{x}_k-\vec{x}^\text{(true)}$
the estimation errors in iteration $k$. After convergence we have in general
$\|\hat{\vec{e}}\|>0$ due to the observation noise. 
The progress of the iterations may thus potentially be judged from several 
different sequences of vectors, e.g.:
\begin{itemize}
\item
the design equation residuals $\vec{s}_k$, whose norm should be minimized;
\item
the vanishing normal equation residuals $\vec{r}_k$;
\item
the vanishing parameter updates $\vec{d}_k=\vec{x}_{k+1}-\vec{x}_k$;
\item
the vanishing truncation errors $\vec{\epsilon}_k$; and
\item
the estimation errors $\vec{e}_k$, which will generally decrease
but not vanish.
\end{itemize}
The last two items are of course not available in the real experiment, but it
may be helpful to study them in simulation experiments. We return in 
Sect.~\ref{sec-conv-crit} to the definition of a convergence criterion 
in terms of the first three sequences.

Given the design matrix $\vec{M}$ and right-hand side $\vec{h}$ (or alternatively
the normals $\vec{N}$, $\vec{b}$), we use the term 
\emph{iteration scheme} for any systematic procedure that generates successive 
approximations $\vec{x}_k$ starting from the arbitrary initial point $\vec{x}_0$ (which
could be zero). The schemes are based on some judicious choice 
of a \emph{preconditioner} matrix $\vec{K}$ that in some sense approximates 
the normal matrix $\vec{N}$ (Sect.~\ref{sec-pre}). The preconditioner must 
be such that the associated system of linear equations, $\vec{K}\vec{x} =\vec{y}$, 
can be solved with relative ease for any $\vec{y}$.

For the astrometric problem $\vec{N}$ is actually 
rank-deficient with a well-defined null space (see Sect.~\ref{sec-back}), and  
we seek in principle the pseudo-inverse solution, 
$\hat{\vec{x}}=\vec{N}^\dagger\vec{b}$, which is orthogonal to the null space. 
By subtracting from each update its projection onto the null space, through the
mechanism described in Sect.~\ref{sec-back}, we ensure that the successive
approximations remain orthogonal to the null space. In this case the circumstance
that the problem is rank-deficient has no impact on the convergence properties
\citep[see][for details]{agis11}.

\subsection{The simple iteration (SI) scheme}
\label{sec-simple}

Given $\vec{N}$, $\vec{b}$, $\vec{K}$ and an initial point $\vec{x}_0$,
successive approximations may be computed as
\begin{equation}\label{eq:si}
\vec{x}_{k+1}= \vec{x}_k + \vec{K}^{-1} \vec{r}_k \, ,
\end{equation}
which is referred to as the \emph{simple iteration} (SI) scheme. Its 
convergence is not guaranteed unless the absolute values of the eigenvalues 
of the so-called iteration matrix $\vec{I}-\vec{K}^{-1}\vec{N}$ are all strictly 
less than one, i.e., $|\lambda_\text{max}|<1$ where $\lambda_\text{max}$
is the eigenvalue with the largest absolute value. In this case it can be 
shown that the ratio of the norms of successive updates asymptotically 
approaches $|\lambda_\text{max}|$. Naturally, $|\lambda_\text{max}|$ will 
depend on the choice of $\vec{K}$. The closer it is to 1, the slower the SI 
scheme converges.

Depending on the choice of the preconditioner, the 
simple iteration scheme may represent some classical iterative solution 
method. For example, if $\vec{K}$ is the diagonal of $\vec{N}$ then the 
scheme is called the Jacobi method; if $\vec{K}$ is the lower triangular 
part of $\vec{N}$ then it is called the Gauss--Seidel method.

\subsection{The conjugate gradient (CG) scheme}
\label{sec-prop-cg}

The normal matrix $\vec{N}$ defines the metric of a scalar product in 
the space of unknowns $\mathbb{R}^n$. Two non-zero vectors $\vec{u}$,
$\vec{v}\in\mathbb{R}^n$ are said to be conjugate in this metric if 
$\vec{u}'\vec{N}\vec{v}=0$. It is possible to find $n$ non-zero vectors 
in $\mathbb{R}^n$ that are mutually conjugate. If $\vec{N}$ is positive
definite, these vectors constitute a basis for $\mathbb{R}^n$.

Let $\{\vec{p}_0,\hdots,\vec{p}_{n-1}\}$ be such a conjugate basis. 
The desired solution can be expanded in this basis as
$\hat{\vec{x}}=\vec{x}_0+\sum_{k=0}^{n-1}\alpha_k\vec{p}_k$. Mathematically,
the sequence of approximations generated by the CG scheme corresponds
to the truncated expansion
\begin{equation}
\label{eq:CGxexp}
\vec{x}_k = \vec{x}_0+\sum_{\varkappa=0}^{k-1}\alpha_\varkappa\vec{p}_\varkappa \, ,
\end{equation}
with residual vectors
\begin{equation}
\label{eq:CGrexp}
\vec{r}_k \equiv \vec{N} (\hat{\vec{x}} - \vec{x}_k) = 
\sum_{\varkappa=k}^{n-1}\alpha_\varkappa\vec{N}\vec{p}_\varkappa \, .
\end{equation}
Since $\vec{x}_n=\hat{\vec{x}}$ it follows, in principle, that the CG converges to
the exact solution in at most $n$ iterations. This is of little practical use, 
however, since $n$ is a very large number and rounding errors in any case 
will modify the sequence of approximations long before this theoretical point 
is reached. The practical importance of the CG algorithm instead lies in the 
remarkable circumstance that a very good approximation to the exact
solution is usually reached for $k \ll n$. 

From Eq.~(\ref{eq:CGrexp}) it is readily seen that $\vec{r}_k$ is orthogonal 
to each of the basis vectors $\vec{p}_0,\hdots,\vec{p}_{k-1}$, and that 
$\alpha_k={\vec{p}_k}'\vec{r}_k/(\vec{p}_k'\vec{N}\vec{p}_k)$. In the CG scheme a 
conjugate basis is built up, step by step, at the same time as successive 
approximations of the solution are computed. The first basis vector is taken 
to be $\vec{r}_0$, the next one is the conjugate vector closest to the resulting 
$\vec{r}_1$, and so on. 

Using that $\vec{x}_{k+1}=\vec{x}_k+\alpha_k\vec{p}_k$ from Eq.~(\ref{eq:CGxexp}),
we have $\vec{s}_{k+1}=\vec{s}_k-\alpha_k\vec{M}\vec{p}_k$ from which
\begin{equation}
\label{eq:CGssr}
\|\vec{s}_{k+1}\|^2 = \|\vec{s}_k\|^2 - \alpha_k^2\,\vec{p}_k'\vec{N}\vec{p}_k 
\le \|\vec{s}_k\|^2 \, .
\end{equation}
Each iteration of the CG algorithm is therefore expected to decrease the norm
of the \emph{design equation} residuals $\|\vec{s}_k\|$. By contrast, although 
the norm of the \emph{normal equation} residual $\|\vec{r}_k\|$ vanishes for 
sufficiently large $k$, it does not necessarily decrease monotonically, and 
indeed can temporarily increase in some iterations. 

Using the CG in combination with a preconditioner $\vec{K}$ means that
the above scheme is applied to the solution of the pre-conditioned
normal equations
\begin{equation}\label{eq:pne}
\vec{K}^{-1}\vec{N}\vec{x} = \vec{K}^{-1}\vec{b} \, . 
\end{equation}
For non-singular $\vec{K}$ the solution of this system is clearly 
the same as for the original normals in Eq.~(\ref{eq:ne}), i.e., $\hat{\vec{x}}$.  
Using a preconditioner can significantly reduce the number 
of CG iterations needed to reach a good approximation of $\hat{\vec{x}}$. 
In Sect.~\ref{sec-algo} and Appendix~\ref{sec-CG-van} we describe in 
more detail the proposed algorithm, based on \cite{vanVorst:03}.

\subsection{Some possible preconditioners}
\label{sec-pre}

The convergence properties of an iterative scheme such as the CG strongly 
depend on the choice of preconditioner, which is therefore a critical step 
in the construction of the algorithm.  The choice represents a compromise
between the complexity of solving the linear system $\vec{K}\vec{x} = \vec{y}$ and
the proximity of this system to the original one in Eq.~(\ref{eq:ne}).
Considering the sparseness structure of $\vec{M}'\vec{M}$ there are some 
`natural' choices for $\vec{K}$. For the astrometric core solution with only
source and attitude unknowns, the design equations for source 
$i=1\dots p$ (where $p$ is the number of primary sources) can be summarized
\begin{equation}\label{eq:design1}
\vec{S}_i\vec{x}_{si}+\vec{A}_i\vec{x}_{a}=\vec{h}_{si}  \, ,
\end{equation}
with $\vec{x}_{si}$ and $\vec{x}_a$ being the source and attitude parts of the 
unknown parameter vector $\vec{x}$ \citep[for details, see][]{bom-lin-hol-09}.
The normal equations (\ref{eq:ne}) then take the form
\begin{equation}\label{eq:norm1}
    \begin{bmatrix}
         \vec{S}_1{\!'}\vec{S}_1  &    0   & \hdots &
                   0    &  \vec{S}_1{\!'}\vec{A}_1   \\[3pt]
          0   &   \vec{S}_2{\!'}\vec{S}_2  & \hdots &
                   0    &  \vec{S}_2{\!'}\vec{A}_2   \\[3pt]
       \vdots & \vdots & \ddots & \vdots & \vdots \\[3pt]
          0   &    0   & \hdots &  \vec{S}_p{\!'}\vec{S}_p
              &  \vec{S}_p{\!'}\vec{A}_p   \\[3pt]
       \vec{A}_1{\!'}\vec{S}_1 & \vec{A}_2{\!'}\vec{S}_2
              & \hdots & \vec{A}_p{\!'}\vec{S}_p &
              \sum_i \vec{A}_i{\!'}\vec{A}_i \\
    \end{bmatrix}
    \begin{bmatrix}
        \vec{x}_1 \\[3pt]
        \vec{x}_2 \\[3pt]
        \vdots  \\[3pt]
        \vec{x}_p \\[3pt]
        \vec{x}_a     \\
    \end{bmatrix}
=
    \begin{bmatrix}
        \vec{S}_1{\!'}\vec{h}_{s1} \\[3pt]
        \vec{S}_2{\!'}\vec{h}_{s2} \\[3pt]
        \vdots                \\[3pt]
        \vec{S}_p{\!'}\vec{h}_{sp} \\[3pt]
        \sum_i \vec{A}_i{\!'}\vec{h}_{si} \\
    \end{bmatrix}\, .
\end{equation}
It is important to note that the matrices $\vec{N}_{si}\equiv\vec{S}_i{\!'}\vec{S}_i$ 
are small (typically $5\times 5$), and that the matrix 
$\vec{N}_a\equiv\sum_i \vec{A}_i{\!'}\vec{A}_i$, albeit large, has a simple 
band-diagonal structure thanks to our choice of representing the
attitude through short-ranged splines.
Moreover, natural gaps in the observation sequence 
make it possible to break up this last matrix into smaller attitude 
segments (indexed $j$ in the following) resulting in a blockwise 
band-diagonal structure. The band-diagonal block associated with
attitude segment $j$ is denoted $\vec{N}_{aj}$; hence
$\vec{N}_a=\text{diag}(\vec{N}_{a1},\vec{N}_{a2},\dots)$.

Considering only the diagonal blocks in the normal matrix, we obtain the 
\emph{block Jacobi preconditioner},
\begin{equation}
\vec{K}_1=
    \begin{bmatrix}
         \vec{S}_1{\!'}\vec{S}_1  &    0   & \hdots &
                   0    &  0   \\[3pt]
          0   &   \vec{S}_2{\!'}\vec{S}_2  & \hdots &
                   0    &  0   \\[3pt]
       \vdots & \vdots & \ddots & \vdots & \vdots \\[3pt]
          0   &    0   & \hdots &  \vec{S}_p{\!'}\vec{S}_p
              &  0   \\[3pt]
       0 & 0 & \hdots & 0 &
              \sum_i \vec{A}_i{\!'}\vec{A}_i \\
    \end{bmatrix} \, .
\end{equation}
Since the diagonal blocks correspond to independent systems that can 
be solved very easily, it is clear that $\vec{K}_1\vec{x}=\vec{y}$ can readily
be solved for any $\vec{y}$.
 
Considering in addition the lower triangular blocks we obtain the 
\emph{block Gauss--Seidel preconditioner},
\begin{equation}
\vec{K}_2=
    \begin{bmatrix}
         \vec{S}_1{\!'}\vec{S}_1  &    0   & \hdots &
                   0    &  0   \\[3pt]
          0   &   \vec{S}_2{\!'}\vec{S}_2  & \hdots &
                   0    &  0   \\[3pt]
       \vdots & \vdots & \ddots & \vdots & \vdots \\[3pt]
          0   &    0   & \hdots &  \vec{S}_p{\!'}\vec{S}_p
              &  0   \\[3pt]
       \vec{A}_1{\!'}\vec{S}_1 & \vec{A}_2{\!'}\vec{S}_2
              & \hdots & \vec{A}_p{\!'}\vec{S}_p &
              \sum_i \vec{A}_i{\!'}\vec{A}_i \\
    \end{bmatrix} \, .
\end{equation}
Again, considering the simple structure of the diagonal blocks, it is 
clear that $\vec{K}_2\vec{x}=\vec{y}$ can be solved for any $\vec{y}$ by first solving
each $\vec{x}_{si}$, whereupon substitution into the last row of equations
allows to solve $\vec{x}_a$.

$\vec{K}_2$ is non-symmetric and it is conceivable that this property
is unfavourable for the convergence of some problems. On the other 
hand, the symmetric $\vec{K}_1$ completely ignores the off-diagonal
blocks in $\vec{N}$, which is clearly undesirable. 
The \emph{symmetric block Gauss--Seidel preconditioner}
\begin{equation}
\label{eq:k3}
\vec{K}_3 = \vec{K}_2 \vec{K}_1^{-1} \vec{K}_2' 
\end{equation}
makes use of the off-diagonal blocks while retaining symmetry. 
The corresponding equations $\vec{K}_3\vec{x}=\vec{y}$ can be solved as two 
successive triangular systems: first, $\vec{K}_2\vec{z}=\vec{y}$ is solved for 
$\vec{z}$, then $\vec{K}_1^{-1}\vec{K}_2'\vec{x}=\vec{z}$ is solved for $\vec{x}$ 
(see below).
It thus comes with the penalty of requiring roughly twice as many
arithmetic operations per iteration as the non-symmetric 
Gauss--Seidel preconditioner.

If the normal matrix in Eq.~(\ref{eq:norm1}) is formally written as
\begin{equation}\label{eq:norm1a}
    \vec{N} =
    \begin{bmatrix}
         \vec{N}_s  &   \vec{L}'   \\
         \vec{L}      &   \vec{N}_a\\
    \end{bmatrix},
\end{equation}
where
$\vec{L}$ is the block-triangular matrix below the main diagonal, and 
$\vec{N}_a=\sum_i \vec{A}_i{\!'}\vec{A}_i$, the preconditioners become
\begin{multline}\label{eq:Kform}
    \vec{K}_1 =
    \begin{bmatrix}
         \vec{N}_s  &   \vec{0}   \\
         \vec{0}      &   \vec{N}_a\\
    \end{bmatrix}, ~
    \vec{K}_2 =
    \begin{bmatrix}
         \vec{N}_s  &   \vec{0}   \\
         \vec{L}      &   \vec{N}_a\\
    \end{bmatrix}, ~
    \vec{K}_3 =
    \begin{bmatrix}
         \vec{N}_s  &   \vec{L}'   \\
         \vec{L}      &   \vec{N}_a\!+\!\vec{L}\vec{N}_s^{-1}\vec{L}'\\
    \end{bmatrix}\\
    \dots
\end{multline}
The second system to be solved for the symmetric block Gauss--Seidel 
preconditioner involves the matrix
\begin{equation}\label{eq:KKK}
    \vec{K}_1^{-1}\vec{K}_2' =
    \begin{bmatrix}
         \vec{I}  &   \vec{N}_s^{-1}\vec{L}'  \\
         \vec{0}      &   \vec{I}\\
    \end{bmatrix},
\end{equation}
where $\vec{I}$ is the identity matrix. This second step therefore 
does not affect the attitude part of the solution vector.

\section{Algorithms}
\label{sec-algo}

In this section we present in pseudo-code some algorithms that implement 
the astrometric core solution using SI or CG. They are described in some 
detail since, despite being derived from well-known classical methods, they 
have to operate within an existing framework (viz., AGIS) which 
allows to handle the very large number of unknowns and observations in 
an efficient manner. Indeed, the numerical behaviour of an algorithm may
depend significantly on implementation details such as the order of certain
operations, even if they are mathematically equivalent. 

In the following, we distinguish between the already introduced
\emph{iterative schemes} on one hand, and the \emph{kernels} on
the other. The kernels are designed to set up and solve the preconditioner 
equations, and therefore encapsulate the computationally complex 
matrix--vector operations of each iteration. By contrast, the iteration schemes
typically involve only scalar and vector operations.  The AGIS framework has 
been set up to perform (as one of its tasks) a particular type of kernel operation,
and it has been demonstrated that this can be done efficiently for the full-size
astrometric problem \citep{Lammers+09}. By formulating the CG algorithm in 
terms of identical or similar kernel operations, it is likely that it, too, can be 
efficiently implemented with only minor changes to the AGIS framework.

The complete solution algorithm is made up of a particular combination of 
kernel and iterative scheme. Each combination has its own convergence 
behaviour, and in Sect.~\ref{sec-conv} we examine some of them.
Although we describe, and have in fact implemented, several different
kernels, most of the subsequent studies focus on the Gauss--Seidel
preconditioner, which turns out to be both simple and efficient.

In the astrometric least-squares problem, the design matrix $\vec{M}$
and the right-hand side $\vec{h}$ of the design equations depend on the current
values of the source and attitude parameters (which together form the vector
of unknowns $\vec{x}$), on the partial derivatives of the observed quantities 
with respect to $\vec{x}$, and on the formal standard error of each observation
(which is used for the weight normalization). Each observation corresponds
to a row of elements in $\vec{M}$ and $\vec{h}$. For practical reasons, these
elements are not stored but recomputed as they are needed, and we may 
generally consider them to be functions of $\vec{x}$. For a particular choice 
of preconditioner and a given $\vec{x}$, the kernel computes the scalar $Q$
and the two vectors $\vec{r}$ and $\vec{w}$ given by
\begin{equation}
\label{eq-kernel}
\left. \begin{aligned}
Q &= \|\vec{h} - \vec{M} \vec{x}\|^2 \, ,\\
\vec{r} & = \vec{M}'(\vec{h} - \vec{M} \vec{x}) \, , \\
\vec{w} & = \vec{K}^{-1} \vec{r} \, .
\end{aligned} \quad \right\} 
\end{equation}
For brevity, this operation is written
\begin{equation}
(Q,\vec{r},\vec{w})\leftarrow\textup{kernel}(\vec{x}) \, .
\end{equation}
For given $\vec{x}$, the vector $\vec{r}$ is thus the right-hand side of normal
equations and $\vec{w}$ is the update suggested by the pre-conditioner,
cf.\ Eq.~(\ref{eq:si}). $Q=\|\vec{s}\|^2$, the sum of the squares of the
design equation residuals, is the $\chi^2$-type quantity to be minimized 
by the least-squares solution; it is needed for monitoring purposes 
(Sect.~\ref{sec-conv-crit}) and should be calculated in the kernel as 
this requires access to the individual observations.  
It can be noted that $\vec{K}$ also depends on $\vec{x}$, although in the linear 
regime (which we assume) this dependence is negligible.

\subsection{Kernel schemes}
\label{sec-algo-kernel}

We have implemented the three preconditioners discussed in Sect.~\ref{sec-pre}, 
viz., the block Jacobi (Algorithm~\ref{algo:kernelJinCG1}), the block Gauss--Seidel
(Algorithm~\ref{algo:kernelGSinCG1}) and the symmetric block Gauss--Seidel
preconditioner (Algorithm~\ref{algo:kernelSGSinCG1}). For the sake of simplicity, 
the algorithms presented here considers only the source and attitude unknowns; 
for the actual data processing they must be extended to include the calibration 
and global parameters as well \citep{agis11}.

In the following, we use $[\vec{B}~|~\vec{b}~\vec{c}~\dots]$ to designate 
a system of equations with  coefficient matrix $\vec{B}$ and right-hand sides 
$\vec{b}$, $\vec{c}$, etc. This notation allows to write compactly several steps
where the coefficient matrix and (one or several) right-hand sides can formally 
be treated as a single matrix. Naturally, the actual coding of the algorithms can 
sometimes also benefit from this compactness. For square, non-singular 
$\vec{B}$ the process of solving the system $\vec{B}\vec{x}=\vec{b}$ is written 
in pseudo-code as $\vec{x}\leftarrow\text{solve}([\vec{B}~|~\vec{b}])$.

A key part of the AGIS framework is the ability to take all the observations 
belonging to a given set of sources and efficiently calculate the corresponding
design equations (\ref{eq:design1}). For each observation $l$ of source $i$, the 
corresponding row of the design equations can be is written 
\begin{equation}
\vec{S}_l\vec{x}_{si}+\vec{A}_l\vec{x}_{aj}=\vec{h}_l \, ,
\end{equation}
where $j$ is the attitude segment to which the observation belongs, $\vec{S}_l$ 
and $\vec{A}_l$ contain the matrix elements associated with the source and attitude
unknowns $\vec{x}_{si}$ and $\vec{x}_{aj}$, respectively.%
\footnote{The observations are normally one-dimensional, in which case 
$\vec{S}_l$ and $\vec{A}_l$ consist of a single row, and the right-hand side 
$\vec{h}_l$ is a scalar.}
In practice, the right-hand side $\vec{h}_l$ for observation $l$ is not a fixed 
number, but is dynamically computed for current parameter values as the 
difference between the observed and calculated quantity, divided by its formal 
standard error. This means that $\vec{h}_l$ takes the place of the design 
equation residual $\vec{s}_l$, and that the resulting $\vec{x}$ must be 
interpreted as a correction to the current parameter values. 
In Algorithms~\ref{algo:kernelJinCG1}--\ref{algo:kernelSGSinCG1} this 
complex set of operations is captured by the pseudo-code statement 
`calculate $\vec{S}_l$, $\vec{A}_l$, $\vec{h}_l$'.

\begin{algorithm}[t]
\caption{-- Kernel scheme with block Jacobi preconditioner}
\label{algo:kernelJinCG1}
\begin{algorithmic}[1]
\STATE{$Q \leftarrow 0$}
\STATE{{\bf for all} attitude segments $j$, zero $[\vec{N}_{aj}~|~\vec{r}_{aj}]$}
\FORALL{sources $i$}
	\STATE{zero $[\vec{N}_{si}~|~\vec{r}_{si}]$}
	\FORALL{observations $l$ of the source}
		\STATE{calculate $\vec{S}_l$, $\vec{A}_l$, $\vec{h}_l$}
		\STATE{$Q \leftarrow Q + \vec{h}_l{\!'}\vec{h}_l$}
		\STATE{$[\vec{N}_{si}~|~\vec{r}_{si}]  \leftarrow [\vec{N}_{si}~|~\vec{r}_{si}]
		+\vec{S}_l{\!'}[\vec{S}_l~|~\vec{h}_l]$}
		\STATE{$[\vec{N}_{aj}~|~\vec{r}_{aj}] \leftarrow [\vec{N}_{aj}~|~\vec{r}_{aj}]
		+\vec{A}_l{\!'}[\vec{A}_l~|~\vec{h}_l]$}
	\ENDFOR
	\STATE{$\vec{w}_{si} \leftarrow\text{solve}([\vec{N}_{si}~|~\vec{r}_{si}])$}
\ENDFOR
\FORALL{attitude segments $j$}
	\STATE{$\vec{w}_{aj}\leftarrow\text{solve}([\vec{N}_{aj}~|~\vec{r}_{aj}])$}
\ENDFOR
\RETURN $Q$, $\vec{r}=(\vec{r}_{s1},\dots,\vec{r}_{a1},\dots)$ and 
$\vec{w}=(\vec{w}_{s1},\dots,\vec{w}_{a1},\dots)$
\end{algorithmic}
\end{algorithm}

\begin{algorithm}[t]
\caption{-- Kernel scheme with block Gauss--Seidel preconditioner}
\label{algo:kernelGSinCG1}
\begin{algorithmic}[1]
\STATE{$Q \leftarrow 0$}
\STATE{{\bf for all} attitude segments $j$, zero $[\vec{N}_{aj}~|~\vec{r}_{aj}]$}
\FORALL{sources $i$}
	\STATE{zero $[\vec{N}_{si}~|~\vec{r}_{si}]$}
	\FORALL{observations $l$ of the source}
		\STATE{calculate $\vec{S}_l$, $\vec{A}_l$, $\vec{h}_l$}
		\STATE{$Q \leftarrow Q + \vec{h}_l{\!'}\vec{h}_l$}
		\STATE{$[\vec{N}_{si}~|~\vec{r}_{si}] \leftarrow [\vec{N}_{si}~|~\vec{r}_{si}]
		+\vec{S}_l{\!'}[\vec{S}_l~|~\vec{h}_l]$}
	\ENDFOR
	\STATE{$\vec{w}_{si} \leftarrow\text{solve}([\vec{N}_{si}~|~\vec{r}_{si}])$}
	\STATE{$\bar{\vec{h}}_{si} \leftarrow \vec{h}_{si}-\vec{S}_i\vec{w}_{si}$}\label{alg2-1}
	\FORALL{observations $l$ of the source}\label{alg-gsk-10}
		\STATE{$[\vec{N}_{aj}~|~\bar{\vec{r}}_{aj}~\vec{r}_{aj}]	\leftarrow 
		[\vec{N}_{aj}~|~\bar{\vec{r}}_{aj}~\vec{r}_{aj}]
		+ 	\vec{A}_l{\!'}[\vec{A}_l~|~\bar{\vec{h}}_l~\vec{h}_l]$}\label{alg2-2}\label{alg-gsk-11}
	\ENDFOR\label{alg-gsk-12}
\ENDFOR
\FORALL{attitude segments $j$}
	\STATE{$\vec{w}_{aj}\leftarrow\text{solve}([\vec{N}_{aj}~|~\bar{\vec{r}}_{aj}])$}
\ENDFOR
\RETURN $Q$, $\vec{r}=(\vec{r}_{s1},\dots,\vec{r}_{a1},\dots)$ and 
$\vec{w}=(\vec{w}_{s1},\dots,\vec{w}_{a1},\dots)$
\end{algorithmic}
\end{algorithm}

\begin{algorithm}[t]
\caption{-- Kernel scheme with symmetric block Gauss--Seidel preconditioner}
\label{algo:kernelSGSinCG1}
\begin{algorithmic}[1]
\STATE{$(Q,\vec{r},\vec{w}) \leftarrow$ kernel$(\vec{x})$ (Algorithm~\ref{algo:kernelGSinCG1})}\label{alg3-2}
\FORALL{sources $i$}
	\STATE{zero $[\vec{N}_{si}~|~\vec{u}_{i}]$}
	\FORALL{observations $l$ of the source}
		\STATE{calculate $\vec{S}_l$, $\vec{A}_l$}
		\STATE{$[\vec{N}_{si}~|~\vec{u}_{i}] \leftarrow [\vec{N}_{si}~|~\vec{u}_i]
		+\vec{S}_l{\!'}[\vec{S}_l~|~(\vec{A}_l\vec{w}_{aj})]$}
	\ENDFOR
	\STATE{$\vec{w}_{si} \leftarrow \vec{w}_{si} - \text{solve}([\vec{N}_{si}~|~\vec{u}_i])$}\label{alg3-1}
\ENDFOR
\RETURN $Q$, $\vec{r}$ and $\vec{w}=(\vec{w}_{s1},\dots,\vec{w}_{a1},\dots)$
\end{algorithmic}
\end{algorithm}

In the block Jacobi kernel (Algorithm~\ref{algo:kernelJinCG1}),
$[\vec{N}_{si}~|~\vec{r}_{si}] \equiv [\vec{S}_i{\!'}\vec{S}_i~|~\vec{S}_i{\!'}\vec{h}_i]$ 
are the systems obtained by disregarding the off-diagonal blocks in the
upper part of Eq.~(\ref{eq:norm1}). Similarly $[\vec{N}_{aj}~|~\vec{r}_{aj}]$, 
for the different attitude segments $j$, together make up the band-diagonal 
system $[\sum_i\vec{A}_i{\!'}\vec{A}_i~|~\sum_i\vec{A}_i{\!'}\vec{h}_i]$ in the
last row of Eq.~(\ref{eq:norm1}).

The kernel scheme for the block Gauss--Seidel preconditioner 
(Algorithm~\ref{algo:kernelGSinCG1}) differs from the above mainly in
that the right-hand sides of the observation equations ($\vec{h}_l$) are 
modified (in line~\ref{alg2-1}) to take into account the change in the source 
parameters, before the normal equations for the attitude segments are 
accumulated. However, since the kernel must also return the right-hand 
side of the normal equations \emph{before} the solution, the original vectors 
$\vec{r}_{aj}$ are carried along in line~\ref{alg2-2}. 

The kernel scheme for the symmetric block Gauss--Seidel preconditioner 
(Algorithm~\ref{algo:kernelSGSinCG1}) is in its first part identical to the
non-symmetric Gauss--Seidel (line~\ref{alg3-2}), but then requires an 
additional pass through all the sources and observations. This second 
pass solves a triangular system with the matrix $\vec{K}_1^{-1}\vec{K}_2'$ given 
in Eq.~(\ref{eq:KKK}). The resulting modification of the source part of 
$\vec{w}$ is done in line~\ref{alg3-1} of Algorithm~\ref{algo:kernelSGSinCG1}. 
Since the design equations are not stored, this second pass through the 
sources and observations roughly doubles the number of calculations 
compared with the non-symmetric Gauss--Seidel kernel.

\subsection{Iteration schemes}
\label{sec-algo-iter}

Comparing Eqs.~(\ref{eq-kernel}) and (\ref{eq:si}) we see that the simple iteration
scheme is just the repeated application of the kernel operation on each approximation,
followed by an update of the approximation by $\vec{w}$. This results in
Algorithm~\ref{algo:simpleItInCG1} for the SI scheme. 
The initialisation of $\vec{x}$ in line~\ref{alg4-1} is arbitrary, as long as it
is within the linear regime -- for example $\vec{x}=\vec{0}$ would do.
The condition in line~\ref{alg4-2} of course needs further specification;
we return to this question in Sect.~\ref{sec-conv}.

\begin{algorithm}[t]
\caption{-- Simple iterative scheme}
\label{algo:simpleItInCG1}
\begin{algorithmic}[1]
\STATE{initialise $\vec{x}$}\label{alg4-1}
\WHILE{$\vec{x}$ not accurate enough}\label{alg4-2}
	\STATE{$(Q,\vec{r},\vec{w}) \leftarrow$ kernel$(\vec{x})$}
	\STATE{$\vec{x} \leftarrow \vec{x} + \vec{w}$}
\ENDWHILE
\end{algorithmic}
\end{algorithm}

The CG scheme (Algorithm~\ref{algo:CGinCG1}) is a particular
implementation of the classical conjugate gradient algorithm with 
preconditioner, derived from the algorithm described in \citet{vanVorst:03} 
as detailed in Appendix~\ref{sec-CG-van}. Whereas most classical algorithms, such 
as the one in \citet{vanVorst:03}, require the multiplication of the normal 
matrix with some vector in addition to the kernel operations involving the 
preconditioner, this specific implementation requires only one
matrix--vector operation per iteration, namely the kernel. This feature
is important in order to allow straightforward implementation in the AGIS 
framework. Indeed, in this form the CG algorithm does not differ significantly 
in complexity from the SI algorithm: the two schemes require about the same
amount of computation and input/output operations per iteration. The
main added complexity is the need to handle three more vectors of length 
$n$ (the total number of unknowns), namely $\vec{p}$ the conjugate direction, 
and $\tilde{\vec{r}}$, $\tilde{\vec{w}}$ to store some intermediate quantities.

\begin{algorithm}[t]
\caption{-- Conjugate gradient scheme}
\label{algo:CGinCG1}
\begin{algorithmic}[1]
\STATE{initialise $\vec{x}$}
\STATE{$(Q,\vec{r},\vec{w}) \leftarrow$ kernel$(\vec{x})$}\label{alg5-2}
\STATE{$\rho \leftarrow \vec{r}\,'\vec{w}$}
\STATE{$\vec{p} \leftarrow \vec{w}$}
\WHILE{ $\vec{x}$ not accurate enough}
	\STATE{$\vec{x} \leftarrow \vec{x} + \vec{p}$}\label{alg5-6}
	\STATE{$(\tilde{Q},\tilde{\vec{r}},\tilde{\vec{w}}) \leftarrow$ kernel$(\vec{x})$}\label{alg5-7}
	\STATE{$\alpha \leftarrow \rho/(\vec{p}\,'(\vec{r}-\tilde{\vec{r}}))$}\label{alg5-8}
	\STATE{$\vec{x} \leftarrow \vec{x} + (\alpha-1)\vec{p}$}\label{alg5-9}
	\STATE{$Q \leftarrow \tilde{Q} - (1-\alpha )^2\rho/\alpha$}\label{alg5-10}
	\STATE{$\vec{r} \leftarrow (1-\alpha ) \vec{r} + \alpha \tilde{\vec{r}}$}\label{alg5-11}
	\STATE{$\vec{w} \leftarrow (1-\alpha ) \vec{w} + \alpha \tilde{\vec{w}}$}\label{alg5-12}
	\STATE{$\rho_\text{old} \leftarrow \rho$}
	\STATE{$\rho \leftarrow \vec{r}\,'\vec{w}$}
	\STATE{$\beta \leftarrow \rho/\rho_\text{old}$}
	\STATE{$\vec{p}\leftarrow \vec{w} + \beta \vec{p}$}
\ENDWHILE
\end{algorithmic}
\end{algorithm}

As explained in Sect.~\ref{sec-prop-cg} the conjugate gradient algorithm
tries to compute a new descent direction conjugate to the previous ones.
In Algorithm~\ref{algo:CGinCG1} the information available to do this 
computation is limited to a few scalars and vectors updated in each
iteration. Hence, if the norm of the design equation residuals fails to 
decrease in an iteration, it could mean that the algorithm was not able
to compute correctly a direction conjugate to the previous ones, due to
accumulation of round-off errors from one iteration to the next. In such
a situation the CG algorithm should be reinitialised. It is equivalent to start 
a new process from the last computed approximation. If this condition
occurs repeatedly in subsequent iterations, then the scheme should be 
stopped, since no better approximation to the least-squares solution
can then be computed using this algorithm. The condition for reinitialisation
could be based either on the sequence of $Q_k$ values returned by 
the kernel, or on $q_k$ calculated from Eq.~(\ref{def-q}). We have found
the former method to be more reliable, in spite of the fact that it depends
on the comparison of large quantities ($Q_{k+1}$ and $Q_k$) that
differ only by an extremely small fraction. However, if $Q_k$ starts to
increase, there is no denying that the CG iterations have ceased to 
work, even if $q_k$ remains positive due to rounding errors.

In the numerical experiments described in Sect.~\ref{sec-conv}
a simple reinitialisation strategy has been used: as soon as the new $Q$
value computed in line~\ref{alg5-10} of Algorithm~\ref{algo:CGinCG1} is not
smaller that the previous value, the scheme returns to line~\ref{alg5-2},
effectively performing an SI step in the next iteration and then continuing 
according to the CG scheme. Too frequent activation of this mechanism 
is prevented by requiring a certain minimum number (say, 5) CG steps 
before another reinitialization can possibly be made. Test runs on larger
problems \citep[e.g., the demonstration solution described in][]{agis11}
suggest that it may be expedient to reinitialize the CG algorithm regularly,
say every 20 to 40 iterations.

\subsection{Frame rotation}
\label{sec-back}

The Gaia observations are invariant to a (small) change of the orientation 
and inertial rotation state of the celestial reference system in which the
astrometric parameters and attitude are expressed. As a consequence, the 
normal matrix $\vec{N}$ has a rank defect of dimension six, corresponding
to three components of the spatial orientation and three components of
the inertial spin of the reference frame. Since the preconditioner $\vec{K}$ 
is always non-singular, the SI and CG schemes still work, but the resulting
positions and proper motions are in general expressed in a slightly different
reference frame from the `true' values, and this frame could moreover change
slightly from one iteration to the next \citep[for details, see][]{agis11}. 
In the numerical tests described below the celestial coordinate frame is
re-oriented at the end of each iteration, in such a way that the derived 
positions and proper motions agree, in a least-squares sense, with their 
true values. This is especially important in order to avoid trivial biases
when monitoring the actual errors of the solution. In the actual processing 
of Gaia data, the frame orientation will instead be fixed by reference
to special objects such as quasars. 

\section{Numerical tests}
\label{sec-conv}

Using numerical simulations of the astrometric core solution we aim to
show that the proposed CG algorithm converges efficiently to the
mathematical least-squares solution of the problem, to within numerical
rounding errors. With simulated data we have the advantage of knowing 
the `true' source parameters, and can therefore use the estimation error 
vector $\vec{e}_k$ as one of the diagnostics. With the real measurements, 
this vector is of course not available, and an important task is to define a 
good convergence criterion based on the actually available quantities
(Sect.~\ref{sec-iter}). 

In this section we first describe briefly the software tool, AGISLab, used for the
simulations, then give and discuss the results of several numerical tests
of the SI and CG algorithms; finally, we discuss some possible convergence
criteria.

\subsection{Simulation tools: AGISLab\label{sec-agislab}}

The Astrometric Global Iterative Solution (AGIS) aims to make astrometric 
core solutions with up to some $5\times 10^8$ (primary) sources, based on about
$4\times 10^{11}$ observations, and is therefore built on a software framework 
specially designed to handle very efficiently the corresponding large data 
volumes and systems of equations. Such a complete solution is expected
to take several weeks on the targeted computer system
\citep[see Sect.~7.3 in][]{agis11}. It is possible to
solve smaller problems in AGIS by reducing the number of included sources;
however, even with the minimum number
($\sim\!10^6$, as determined by the need to have a fair number of sources 
within each field of view at any given time) it could take several days to run 
a solution to full convergence on the computer system currently 
in use. The input data for AGIS are normally the output from a preceding
stage, in which higher-level image parameters are derived from the
raw CCD measurement data. In the DPAC simulation
pipeline, raw satellite data are generated by a separate unit and then fed
through the preprocessing stage before entering the AGIS. This complex
system is necessary in order to guarantee that DPAC will be able to cope
with the real satellite data, but it is rather inflexible and unsuitable for 
more extensive experimentation with different algorithms.

We have therefore developed a scaled-down version of AGIS, called AGISLab, which 
allows us to run simulations with considerably less than $10^6$ sources 
in a correspondingly much shorter time. Moreover, the simulation of the
required input data is an integrated part of AGISLab, so that it is for example
very easy to make several runs with different noise realisations but otherwise 
identical conditions. The scaling uses a single parameter $S$ such that 
$S=1$ leads to an astrometric solution that uses approximately the 
current Gaia design and a minimum of $10^6$ primary sources, while 
$S=0.1$ would only use 10\% as many primary sources, etc. For $S<1$ 
it is necessary to modify the Gaia design used in the simulations in order
to preserve certain key quantities such as the mean number of sources in the 
focal plane at any time, the mean number of field transits of a given source 
over the mission, and the mean number of observations per degree of freedom 
of the attitude model. In practice this is done by formally reducing the focal 
length of the astrometric telescope and the spin rate of the satellite by the 
factor $S^{1/2}$, and increasing the time interval between attitude spline 
knots by the factor $S^{-1}$.

All of the simulation experiments reported here were made with a scaling
parameter $S=0.1$, using $10^5$ sources, an astrometric field of
$\simeq 2.1^\circ\times 2.2^\circ$ per viewing direction, a spin rate of
19~arcsec~s$^{-1}$, and a time interval between attitude knots of 300~s,
corresponding to $1.58^\circ$ on the celestial sphere. Experiments using
different values of $S$ show that the convergence behaviour of the investigated 
solution algorithms does not depend strongly on the scaling. For a full-scale
solution the convergence rate is likely to be lower than in the present
experiments, but not by a significant factor.

AGISLab provides all features to generate a set of true parameter values,
including a random distribution of sources on the celestial sphere and the 
true attitude (e.g., following the nominal Gaia scanning law), and hence
the observations obtained by adding a Gaussian random number to the
computed (`true') observation times. It can also generate starting values
for the source and attitude parameters that deviate from the true values
by random and systematic offsets. Having generated the observations,
AGISLab sets up and solves the least-squares problem using some of the
algorithms described in this paper. Finally, AGISLab contains a number
of utilities to generate statistics and graphical output. 

The present simulations, using $10^5$ sources, span a time interval
of 5~years, generating 87\,610\,420 along-scan and 8\,789\,616 
across-scan observations. The number of source parameters is
500\,000 and the number of (free) attitude parameters is 
1\,577\,889; the total number of unknowns is $n=2\,077\,889$ and
the total number of observations $m=96\,400\,036$. The along-scan
observations consist of the precise times when the source images cross 
certain fiducial lines in the focal plane, nominally at the centre of each
CCD; the across-scan observations consist of the transverse angles
of the images as they enter the first CCD. Although the along-scan
observations are times, all residuals are expressed as angles following
the formalism described in \citet{Lindegren10}.
  
When creating the initial source and attitude parameter errors, some 
care must be exercised to avoid that the initial errors are trivially removed by the 
iterations. For example, if one starts with random initial source errors,
but no attitude errors, then already the first step of the simple iteration 
scheme will completely remove the source errors. This trivial
situation can of course be avoided by assuming some random initial
attitude errors as well. However, it is not realistic to assume independent
attitude errors either -- for example by adding white noise to the attitude
spline coefficients. On the contrary, the initial Gaia attitude will have 
severe and strongly correlated errors depending on the very imperfect 
source catalogue used at that stage. Indeed, the challenge of the
astrometric core solution is precisely to remove this correlation as
completely as possible. It is therefore important to start with initial
attitude errors that somehow emulate this situation. We do that by
first adding random (and in some cases systematic) errors to the source
parameters, and then performing an attitude update by applying the 
attitude block of the Jacobi-like preconditioner. The resulting attitude, 
which then contains a strong imprint of the initial source errors, is 
taken as the initial attitude approximation for the iterative solution. 

\begin{figure*}[t!]
\begin{center}
\centerline{
\includegraphics[width=88mm]{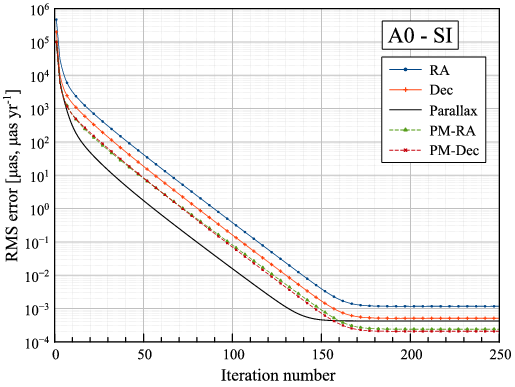}
\hspace{4mm}
\includegraphics[width=88mm]{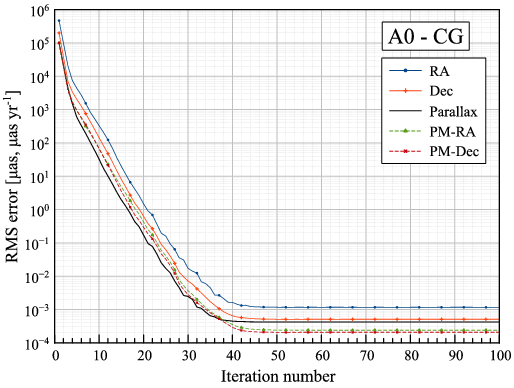}
}\vspace{2mm}
\centerline{
\includegraphics[width=88mm]{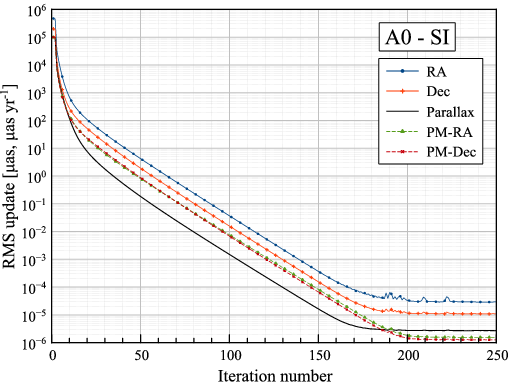}
\hspace{4mm}
\includegraphics[width=88mm]{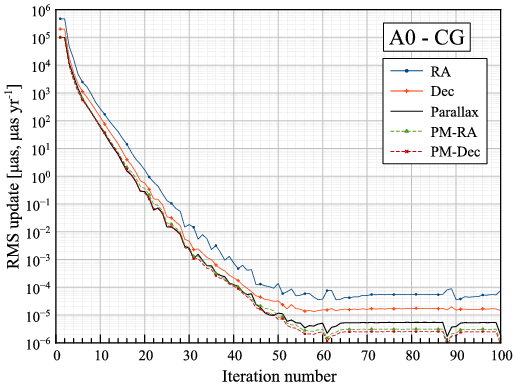}
}
\caption{Convergence plots for test case A0 (without observation noise),
using the simple iteration scheme (SI, left diagrams), and the conjugate
gradient scheme (CG, right diagrams). The top diagrams show the RMS
errors of the astrometric parameter (i.e., the RMS differences between the 
calculated and true values). The bottom diagrams show the RMS updates
of the astrometric parameters.}
\label{fig-conv-A0}
\end{center}
\end{figure*}

To illustrate the convergence of the different iteration schemes, we use 
three kinds of diagrams, which are briefly explained hereafter.

\emph{Convergence plots} show scalar quantities such as the RMS values 
of the errors, updates or residuals, plotted on a logarithmic scale versus 
the iteration number ($k$). For
the error ($\vec{e}_k$) and update ($\vec{d}_k$) vectors we
consider only the source parameters; the attitude errors and updates follow 
similar curves, not adding much information about the convergence behaviour. 
The source parameters are separated according to type ($\alpha*$, $\delta$, 
$\varpi$, $\mu_{\alpha*}$, and $\mu_\delta$).%
\footnote{Following the convention introduced with the Hipparcos and Tycho
Catalogues \citep{esa:97}, we use an asterisk to indicate that differential 
quantities in right ascension include the factor $\cos\delta$ and thus 
represent true (great-circle) angles on the celestial sphere. For example, 
a difference in right ascension is denoted 
$\Delta\alpha*=\Delta\alpha\cos\delta$ and the proper motion 
$\mu_{\alpha*}=(\text{d}\alpha/\text{d}t)\cos\delta$.}
The purpose of these plots is to show the rate of global convergence
of the different algorithms. 

\emph{Error maps} show, for selected iterations, the error in one of the 
astrometric parameters (i.e., its currently estimated value minus the true 
value) as a function of position on the celestial sphere. The purpose of 
these plots is to show graphically the possible
existence of systematic errors in the solution as a function of position on 
the sky. Significant such errors could exist without being noticeable in the 
global convergence plots. To produce these error maps, 
the sky is divided into small bins of roughly equal solid angle, and the 
median error is computed over all
the stars belonging to the bin. A colour is attributed to each bin
according to the median error, and a map is plotted in equatorial 
coordinates, using an equal-area Hammer--Aitoff projection. In the
maps shown, the sky is divided into 12\,288 bins of approximately
$1.8^\circ$ side length; there are on average 8.1~stars per bin. 
We choose to show only the distribution of parallax errors, although
qualitatively similar maps are obtained for each of the five
astrometric parameters. 

\emph{Truncation error maps} are similar to the error maps, but show
the difference between the current iteration and the final (converged)
iteration. They therefore display the type of systematic errors that
could exist in the solution, if the iteration process is prematurely 
terminated. 

\begin{figure*}[t!]
\begin{center}
\centerline{\includegraphics[width=175mm,height=206mm]{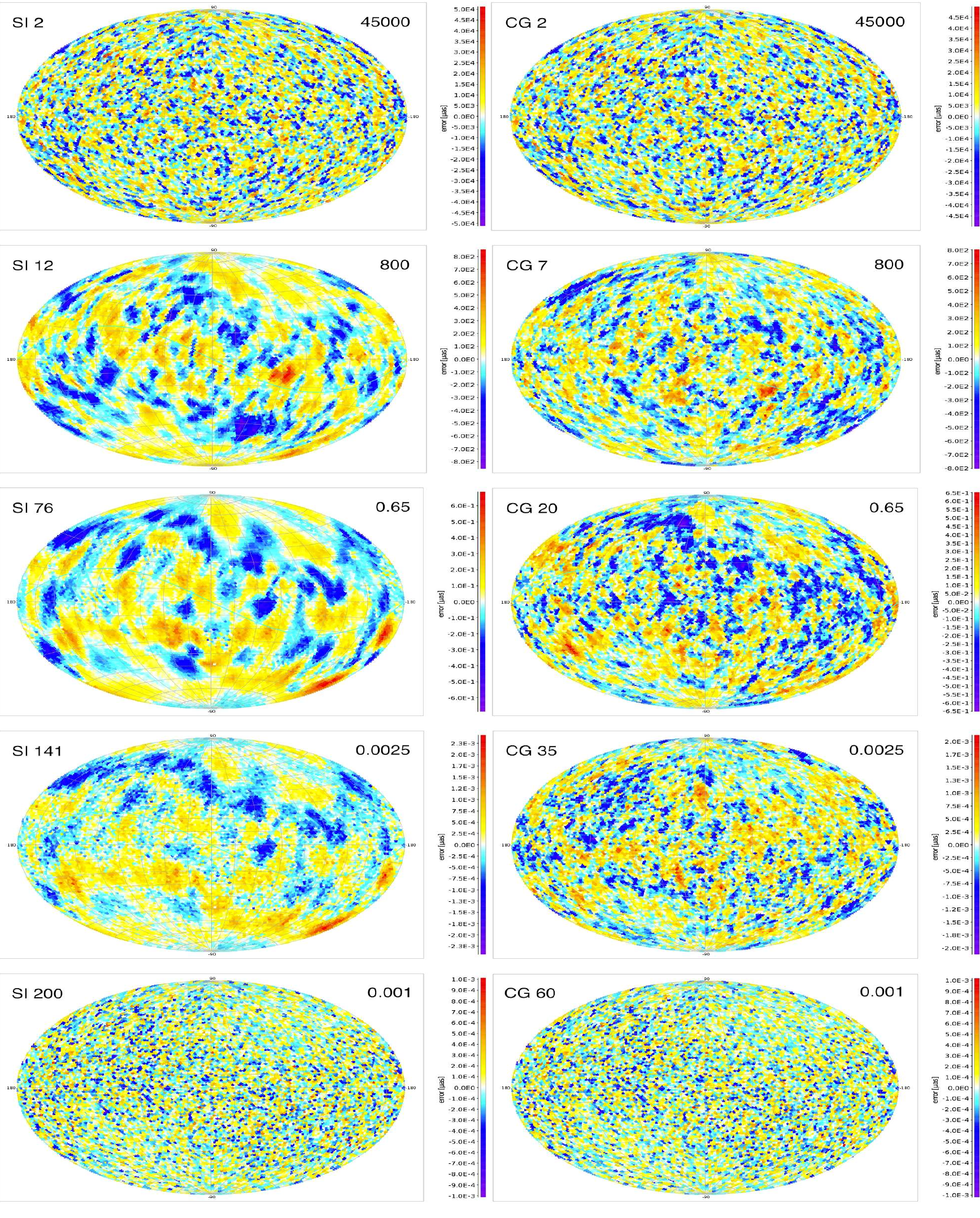}}
\caption{Parallax error maps for test case A0 (no observation noise) at selected points 
in the simple iteration scheme (SI, left column) and conjugate gradient scheme 
(CG, right column). The iteration number is shown in 
the top left corner of each map (the initial values are iteration 2 due to
the `start-up' procedure described in the text). The number in the top right
corner is the approximate amplitude (in $\mu$as) of the colour scale. 
Starting from identical initial values of the astrometric parameters, SI and CG converge 
to the same solution (equal to the true parallaxes to within 
$\pm 0.001~\mu$as), although along different paths; moreover CG 
converges about four times quicker.}
\label{fig-A0}
\end{center}
\end{figure*}

\subsection{Case A: Uniform distribution of sources and weights\label{sec-casea}}

In Case~A we consider a sky of isotropically distributed sources of
uniform brightness, so that they all obtain the same statistical weight
per observation. This weight corresponds to a standard deviation of 
100~$\mu$as for the along-scan (AL) observations and 600~$\mu$as for 
the across-scan (AL) observations. These numbers are representative for Gaia's 
expected performance for bright stars ($G$ magnitude from $\simeq 6$ to 13). 
The top diagrams in Fig.~\ref{fig-A0} show the distribution of initial parallax 
errors on the sky; the amplitude of these errors is about $\pm 45$~mas.

Three separate tests, subsequently denoted A0, A1, and A2, were made with 
the uniform source distribution in Case~A:
\begin{description}
\item[A0:] No observational errors were added to the computed observations.
Consequently both the SI and the CG should converge to the true source parameters.
\item[A1:] Random centred Gaussian errors were added to the computed 
observations, with standard deviations equal to the nominal standard errors 
(100 and 600~$\mu$as AL and AC). Again, both SI and CG should converge to
the same source parameters, which however will differ from the true values by
several $\mu$as due to the observation noise. 
\item[A2:] This test used exactly the same noisy observations as A1, but the
iterations start with a different set of initial values. After convergence, the 
solution should be exactly the same as in A1. This test was only made with 
the CG algorithm.
\end{description}
In all cases the Gauss--Seidel preconditioner (Algorithm~\ref{algo:kernelGSinCG1}) 
was used for both the SI and CG schemes. We have also tested the symmetric 
Gauss--Seidel preconditioner (Algorithm~\ref{algo:kernelSGSinCG1}) on a 
smaller version ($S=0.01$) of this problem, but without any significant 
improvement in the convergence over the (non-symmetric) Gauss--Seidel 
preconditioner.

\begin{figure*}[t!]
\begin{center}
\centerline{
\includegraphics[width=88mm]{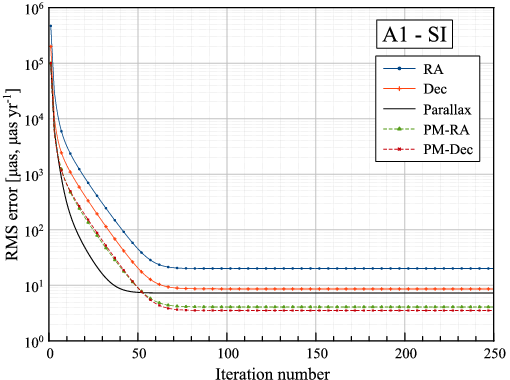}
\hspace{4mm}
\includegraphics[width=88mm]{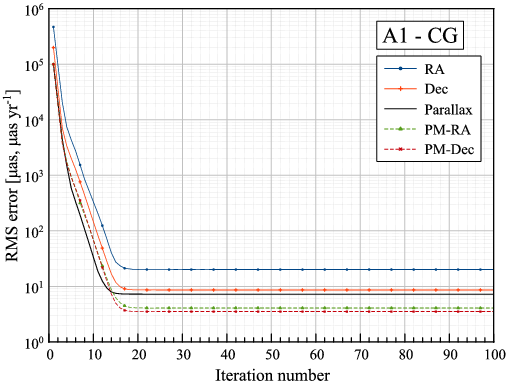}
}\vspace{2mm}
\centerline{
\includegraphics[width=88mm]{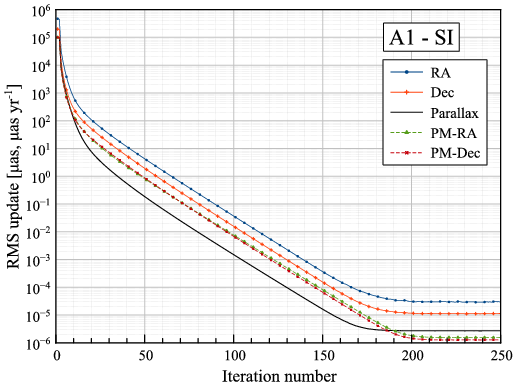}
\hspace{4mm}
\includegraphics[width=88mm]{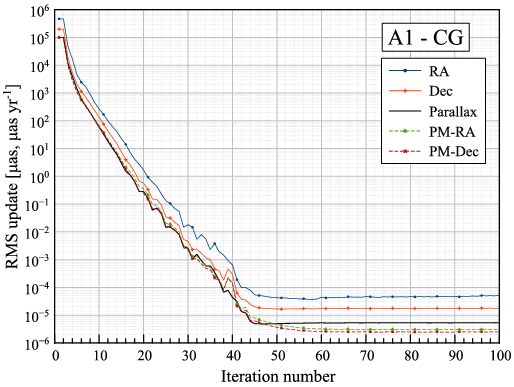}
}
\caption{Convergence plots for test case A1 (including observation noise),
using the simple iteration scheme (SI, left diagrams), and the conjugate
gradient scheme (CG, right diagrams). See Fig.~\ref{fig-conv-A0} for
further explanation.}
\label{fig-conv-A1}
\end{center}
\end{figure*}

\begin{figure*}[t!]
\begin{center}
\centerline{\includegraphics[]{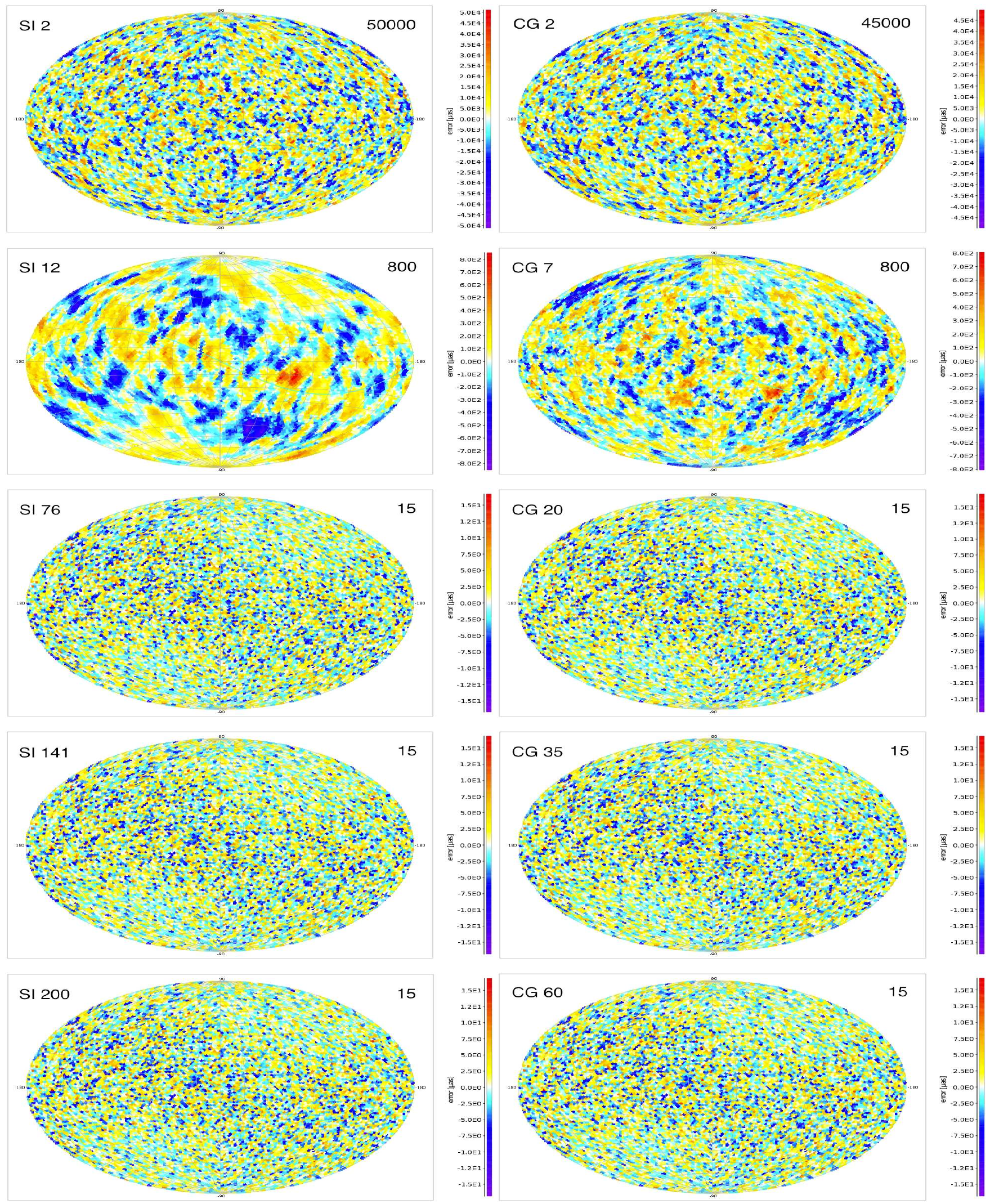}}
\caption{Error maps for test case A1 (including observation noise)
at selected points in the
simple iteration scheme (SI, left column) and conjugate gradient scheme 
(CG, right column). See Fig.~\ref{fig-A0} for further explanation
of the diagram layout. Starting from identical initial parallax values, 
SI and CG converge to the same solution, although along different paths.
Apparently, the parallaxes have converged more or less to their final
values after iteration 76 (SI) or 20 (CG), but as shown in Fig.~\ref{fig-A1a}
there are then still significant, spatially correlated truncation errors that
require many more iterations to be completely removed.}
\label{fig-A1}
\end{center}
\end{figure*}

\begin{figure*}[t!]
\begin{center}
\centerline{\includegraphics[]{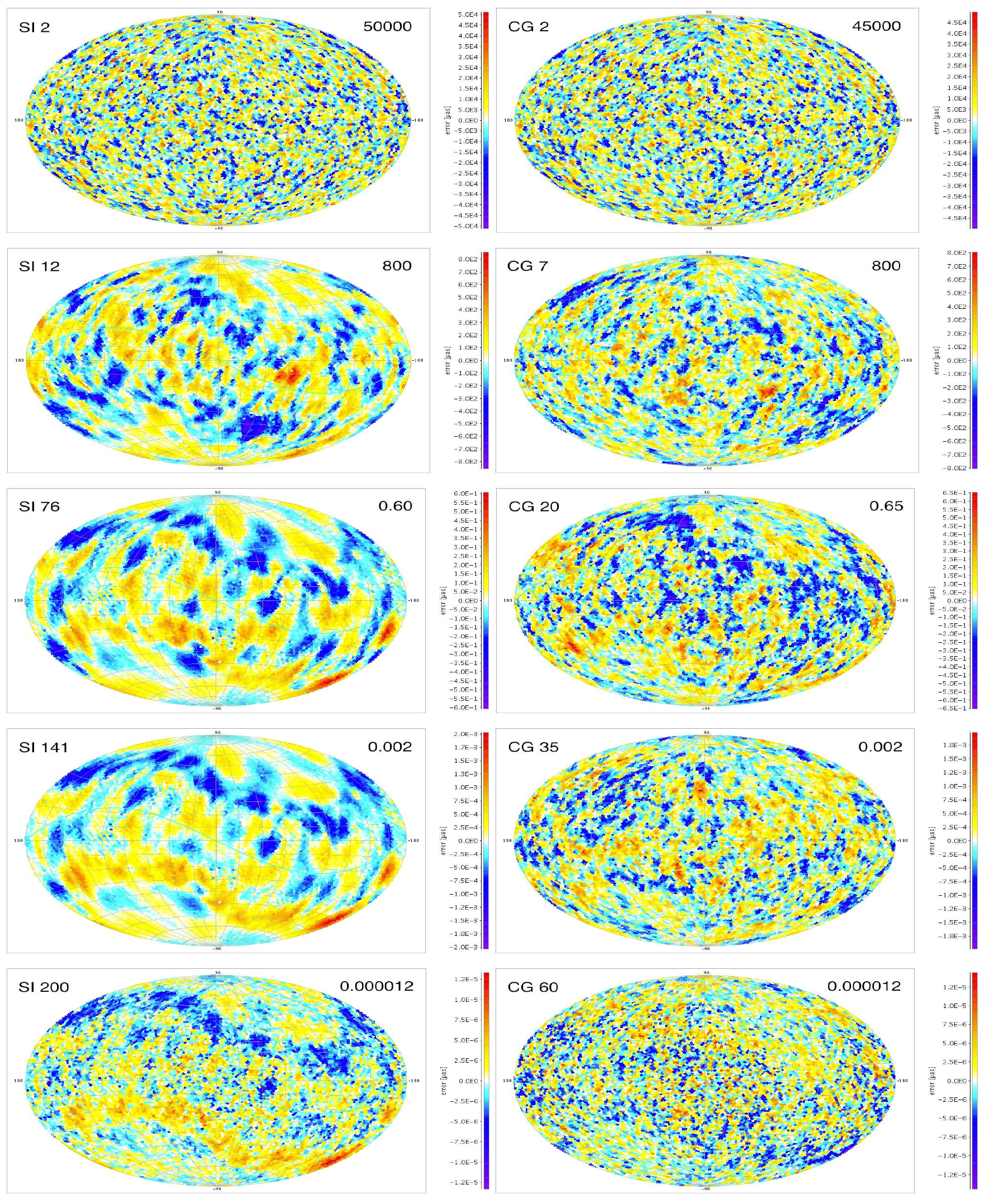}}
\caption{Truncation error maps for test case A1 at selected points
in the simple iteration scheme (SI, left column) and 
conjugate gradient scheme (CG, right column). 
See Fig.~\ref{fig-A0} for further explanation of the diagram layout. 
Each map shows the difference between the result in the current iteration 
and the (presumably converged) solution obtained after 250 (SI) or 
100 (CG) iterations.  
Although the parallaxes have converged more or less to their final
values after iteration 76 (SI) or 20 (CG), as shown in Fig.~\ref{fig-A1},
there are then still spatially correlated truncation errors at the
$\pm 0.6~\mu$as level in both solutions.}
\label{fig-A1a}
\end{center}
\end{figure*}

\subsubsection{Test case A0: Comparing SI and CG without 
noise}\label{sec:A0}

Figure~\ref{fig-conv-A0} shows the global convergence for test case
A0, i.e., without observation noise. The top diagrams show the 
errors of the astrometric parameters, and the bottom diagrams 
the updates. The left diagrams
are for the SI scheme, and the right diagrams for CG.  
The errors and updates are expressed in $\mu$as (for $\alpha*$, 
$\delta$, and $\varpi$) and $\mu$as~yr$^{-1}$ 
(for $\mu_{\alpha*}$ and $\mu_\delta$).

From Fig.~\ref{fig-conv-A0} it is seen that both algorithms eventually 
reach the same level of RMS errors ($\la 0.001~\mu$as in position and
parallax and $\la 0.001~\mu$as~yr$^{-1}$ in proper motion), 
and that the updates settle at levels that are 1--2~dex below the errors. 
The updates do not systematically
decrease beyond iteration $\sim\! 200$ (SI) and $\sim\! 60$ (CG),
suggesting that the full numerical precision has been reached at 
these points.

The maps of parallax errors at selected iterations of test case A0 are shown 
in Fig.~\ref{fig-A0}. The top maps show the initial errors (which are the
same for SI and CG) and the bottom maps show the (apparently) converged 
results in iteration 200 (SI) and 60 (CG), respectively. The selection of 
intermediate results, although somewhat arbitrary, was made at comparable 
levels of truncation errors in the two algorithms. It is noted that CG converges 
three to four times faster than SI, in terms of the number of iterations 
required to reach a given level of truncation errors. Furthermore, it is seen
that the converged error maps look quite identical. Inspection of the 
numerical results shows that this is indeed the case: whereas the RMS
parallax error is $4.24\times 10^{-4}~\mu$as both in SI (iteration 200) 
and CG (iteration 60), the RMS value of the difference between the two 
sets of parallaxes is only $5.33\times 10^{-6}~\mu$as.
This means that both algorithms have found virtually exactly the same 
solution, although one that deviates slightly from the true one. 

The likely cause of this deviation is quantization noise when computing
the observation times in the AGISLab simulations, as shown by the following
considerations. Because double-precision
arithmetic is not accurate enough to represent the observation times over
several years, they are instead expressed in nano-seconds (ns) and stored 
as long (64 bit) integers. In the present simulations, which use a scaling 
factor $S=0.1$ (see Sect.~\ref{sec-agislab}), the satellite spin rate is about 
19~arcsec~s$^{-1}$, and the least significant bit of the stored observation 
times therefore corresponds to $0.019~\mu$as. This generates a (uniformly distributed) 
observation noise with an RMS value of $0.019\times 12^{-1/2}=0.0055~\mu$as. 
In order to estimate the corresponding parallax errors, we note that the
ratio of the RMS parallax errors to the RMS observation noise depends 
only on the mean number of observations per source and on certain 
temporal and geometrical factors related to the scanning law, and is 
therefore invariant to a scaling of the observation noise. The ratio can
be estimated from the A1 tests discussed in Sect.~\ref{sec:A1}, which
use an observation noise of $100~\mu$as and give an RMS parallax error
of $7.26~\mu$as. The expected RMS parallax error due to the quantization 
of the observation times is then $0.0726\times 0.0055=0.00040~\mu$as,
in fair agreement with the parallax errors of the `noiseless' A0 tests.

Returning to the error maps in Fig.~\ref{fig-A0}, a further observation 
is that the SI rather quickly develops a certain error pattern (most
clearly seen in the map designated SI\,76), correlated over some 
10--20$^\circ$, which
only slowly fades away with more iterations, until it completely
disappears. This can be understood in relation to the iteration matrix
mentioned in Sect.~\ref{sec-simple}: the dominant pattern shows the 
eigenvector corresponding to the largest eigenvalue of the iteration
matrix. This is consistent with the very straight lines in the left
diagrams of Fig.~\ref{fig-conv-A0} between iterations $\sim\,$50
and 120, showing a geometric progression with a factor 0.91 
improvement between successive iterations; we interpret this
as $|\lambda_\text{max}|\simeq 0.91$. By contrast, the error maps
for the CG scheme do not exhibit similar persistent patterns,
have a smaller correlation length, and the convergence is only
very roughly geometric and not even monotonic at all times.

\subsubsection{Test case A1: Comparing SI and CG with noise}\label{sec:A1}

Figures~\ref{fig-conv-A1} and \ref{fig-A1} show the corresponding
convergence plots and error maps for test case A1, where the
simulated observations include a nominal noise. The convergence plots
(Fig.~\ref{fig-conv-A1}) show that the RMS errors have already settled in
iteration 70 (SI) or 20 (CG), at which points the solutions are however
far from converged, as shown by the RMS updates in the lower diagrams. 
The full convergence is only reached at iteration
200 (SI) or 60 (CG), exactly as in the noiseless case (A0). The updates then
settle at about the same levels as in case A0. The rate of convergence is 
therefore not significantly affected by the noise (if anything, the noise
seems to have a slightly stabilizing effect in the final iterations before
convergence).

The error maps (Fig.~\ref{fig-A1}) start,  in the top diagrams, at the same 
initial approximation in case A0, and develop along different paths to the
converged solutions in the bottom diagrams, which are virtually identical 
for SI and CG. Inspection of the numerical results confirms that the two 
algorithms have indeed converged to the same solution, within the rounding 
errors: for example, the RMS values of the parallax error in SI (iteration 200) 
and CG (iteration 60) are both $7.26~\mu$as, while the RMS difference 
between the solutions is $5.16\times 10^{-6}~\mu$as.

Although the error maps in Fig.~\ref{fig-A1} do not appear to change 
much after 
iteration 76 (SI) and 20 (CG), we inferred from the convergence plots that
neither solution was truly converged at these points. In order to examine
the evolution of the errors beyond these points, we show in Fig.~\ref{fig-A1a} 
the truncation errors in parallax, i.e., the difference between the solution at
a given iteration and the solution after the maximum number of iterations
(250 for SI and 100 for CG). Interestingly, the truncation error maps in
Fig.~\ref{fig-A1a} look very much like the error maps in Fig.~\ref{fig-A0}
for the noiseless case (A0). The iterations therefore follow more or less the
same path through solution space, independent of the observation noise
(but of course different in SI and CG). This is consistent with our previous
observation that A0 and A1 require the same number of iterations for full
convergence. -- It is noted that the truncation errors for the SI in 
iteration 200 still show a residual pattern clearly related to the scanning 
law (with systematically negative and positive parallax errors around 
ecliptic latitude $+45^\circ$ and $-45^\circ$, respectively), although
the amplitude is very small, about $\pm 5\times 10^{-6}~\mu$as. The
truncation errors for the CG, at iteration 60, have a similar amplitude but
are spatially less correlated.

\begin{figure*}[t!]
\begin{center}
\centerline{\includegraphics[]{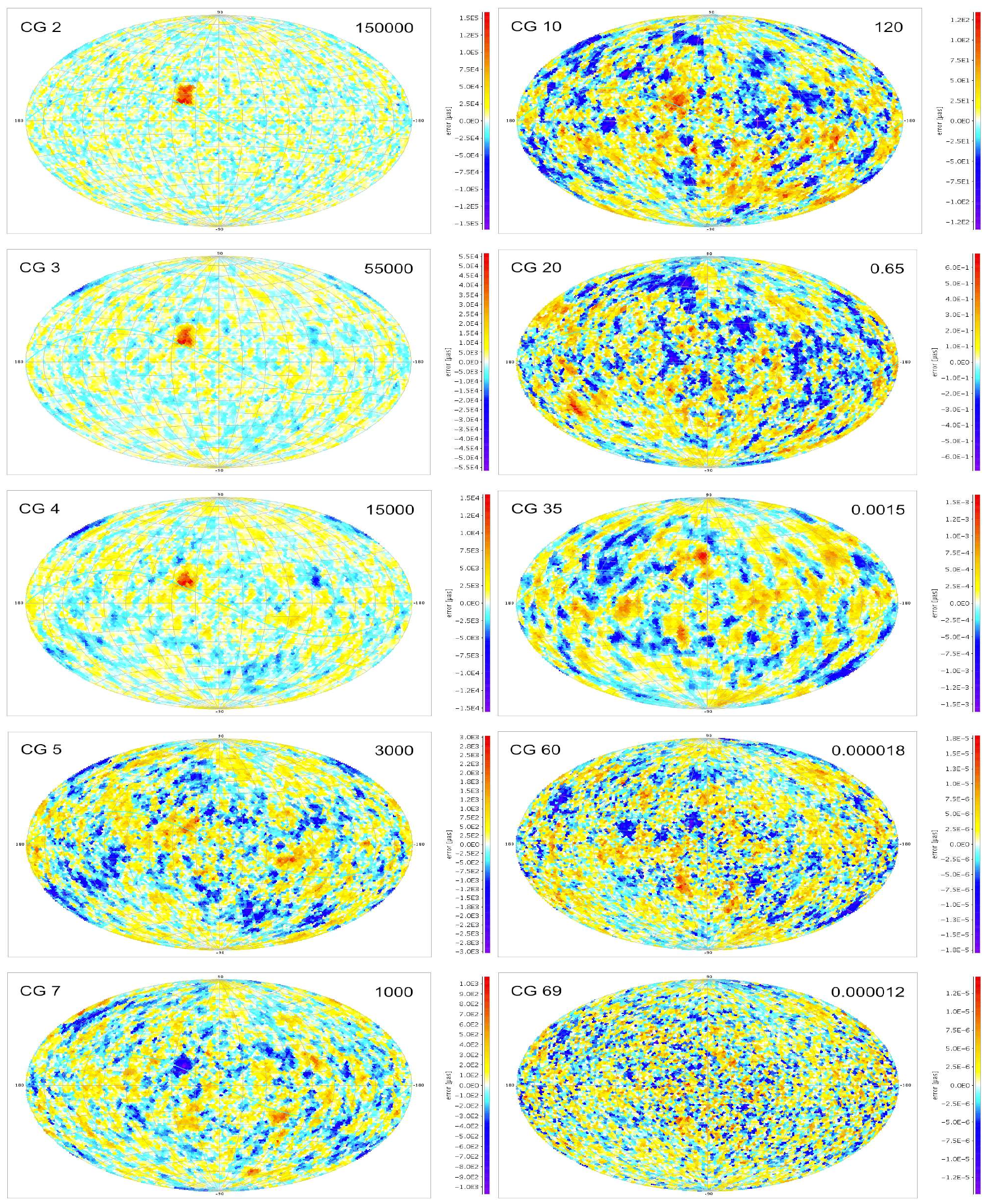}}
\caption{Truncation error maps for test case A2 (same as A1 but with 
different initial values) at selected iterations in the CG scheme. More of 
the first few iterations are shown in order to illustrate the diffusion of the 
large, localized initial errors.  
See Fig.~\ref{fig-A0} for further explanation of the diagram layout.}
\vspace{2mm}
\label{fig-A2}
\end{center}
\end{figure*}

\begin{figure}[t]
\begin{center}
\centerline{
\includegraphics[width=88mm]{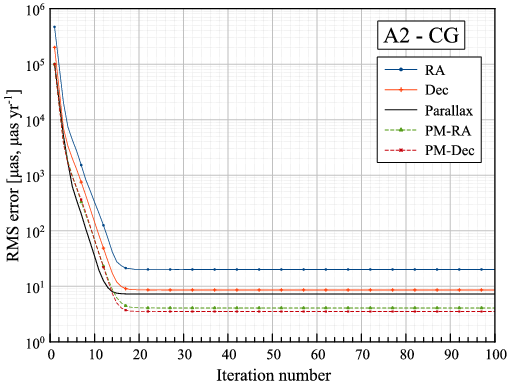}
}\vspace{2mm}
\centerline{
\includegraphics[width=88mm]{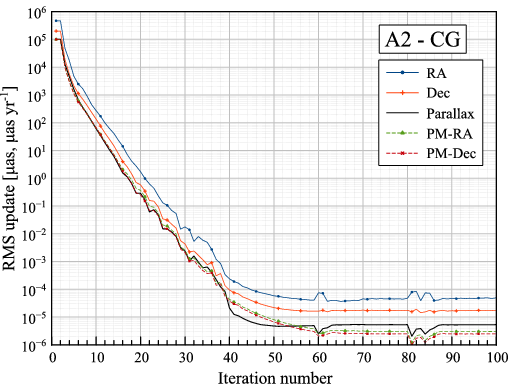}
}
\caption{Convergence plots for test case A2 (same as A1 but with 
different initial values), using the conjugate gradient scheme (CG). 
See Fig.~\ref{fig-conv-A0} for further explanation.}
\label{fig-conv-A2}
\end{center}
\end{figure}

\begin{figure}[t!]
\begin{center}
\centerline{
\includegraphics[width=88mm]{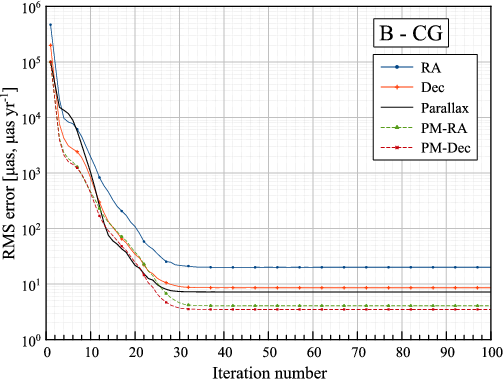}
}\vspace{2mm}
\centerline{
\includegraphics[width=88mm]{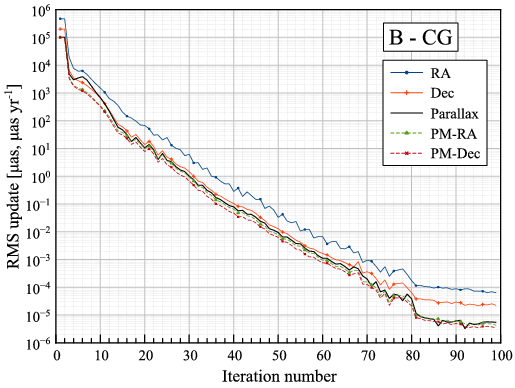}
}
\caption{Convergence plots for test case B (non-uniform weight distribution), 
using the conjugate gradient scheme (CG). 
See Fig.~\ref{fig-conv-A0} for further explanation.}
\label{fig-conv-B}
\end{center}
\end{figure}

\subsubsection{Test case A2: Starting CG from a different point}\label{sec:A2}

The previous tests have all started from the same initial approximation,
illustrated by the errors maps SI2 and CG2 at the top of Figs.~\ref{fig-A0}
and \ref{fig-A1}. The aim of test case A2 is to show that the CG algorithm
finds the same solution, for the same observations as in A1, when starting 
from a different initial approximation. To this end, we added 
0.2~arcsec to the initial parallax values of 504 sources 
in an area of about $200~\text{deg}^2$ centred on $\alpha=30^\circ$,
$\delta={+}20^\circ$; subsequently we refer to this as the `W area'. 
Having strongly deviating initial values in a relatively small area
makes it easy to follow their diffusion among the sources 
in subsequent iterations, e.g., by visual inspection of the error maps
(Fig.~\ref{fig-A2}). A position close to the ecliptic was chosen for the
W area, since the ecliptic region is less well observed by Gaia, due to 
its scanning law, than other parts of the celestial sphere. Potentially, 
therefore, the astrometric solution might be less efficient in eliminating 
the initial errors in such an area.

The convergence plots in Fig.~\ref{fig-conv-A2} are not drastically
different from the corresponding plots in test case A1 (right panels of 
Fig.~\ref{fig-conv-A1}), although the updates do not reduce quite as
quickly after iteration $\sim\,$40. The truncation error maps in
Fig.~\ref{fig-A2} show that the large initial errors in the W area are 
quickly damped
in the first few iterations, and even reversing the sign around iteration
7; after iteration 10 the W area does not stand out. The
subsequent truncation errors maps (e.g., in iteration 20 and 35) are 
remarkably similar to those in test case A1 (right panels of Fig.~\ref{fig-A1}). 
However, the residual large-scale truncation patterns do not completely
disappear, to the same level as in A1, until around iteration 69. 

Numerically, the RMS parallax difference between the A2 and A1
solutions is $4.95\times 10^{-6}~\mu$as in iteration 60, and 
$4.64\times 10^{-6}~\mu$as in iteration 69. Comparing only the
parallaxes for the 504 sources in the W area, the RMS difference is
$5.74\times 10^{-6}~\mu$as. The converged results are thus virtually 
identical; in particular the initial offsets in the W area have been 
reduced by more than 10 orders of magnitude.

\subsection{Case B: Non-uniform distribution of weights}
\label{sec-case-B}

The uniform sky considered in Case~A is highly idealised: the real sky 
contains a very non-uniform distribution of stars of different magnitudes. 
The standard deviation of the along-scan observation noise, $\sigma$, 
is mainly a function of stellar magnitude, and could vary by more than 
a factor 50 between the bright and faint sources. As a result, the 
statistical weight of the observations (defined as the sum of $\sigma^{-2}$ 
for the observations collected in a time interval of a few seconds) is often 
very different in the two fields of view. This happens, for example, when 
one field of view is near the galactic plane and the other is at a high galactic 
latitude, or when a rich and bright stellar cluster passes through one of 
the fields. At such times the along-scan attitude is almost entirely
determined by the observations in the field with the higher weight. 
Intuitively it would seem that this could weaken the connectivity between 
the fields, and consequently the quality of the astrometric solution. 
A particular concern could be that the accuracy of the absolute parallaxes 
of the cluster stars, and their connection to the global reference frame, 
might suffer, since both these qualities critically depend on 
Gaia's ability to measure long arcs by connecting observations in the two 
fields of view. In the new reduction of the Hipparcos data by \citet{vanLee:2007}
special attention was given to the weight distribution between the two
fields of view when performing the attitude solution. As described in 
Sect.~10.5.3 of \citet{vanLee:2007}, the weight ratio was not allowed
to exceed a certain factor ($\sim$3); this was achieved by reducing, 
when necessary, the weights of the observations in one of the fields.
On the other hand, from a more theoretical standpoint it can be argued that 
the intentional removal or down-weighting of perfectly good data cannot 
possibly improve the results.

In order to investigate the impact of an inhomogeneous weight distribution 
on the solution, we present in Case~B a sky with a strong and 
evident contrast in statistical weight. The same source distribution and
initial values were used as in Case~A2, but all the errors of the
observations, as well as their assumed standard errors in the solution, 
were reduced by a factor 5 for the 504 sources in the W area (centred on 
$\alpha=30^\circ$, $\delta ={+}20^\circ$). As before, the starting
values of the parallaxes in the W area were also offset by 0.2~arcsec. 
This case could 
represent a situation where the stars in a single bright cluster obtain 
very accurate individual astrometric measurements, while the initial parallax 
knowledge of the cluster is strongly biased. In Case~B we test the 
ability of the astrometric solution to produce unbiased parallax estimates 
for the cluster, as well as for the rest of the sky, without using any 
weight-balancing schemes such as outlined above. The observations
are strictly weighted by $\sigma^{-2}$, so the weight contrast between
the fields of view is roughly a factor 25 whenever the cluster stars are
observed, while it is about 1 at all other times. 

The convergence plots in Fig.~\ref{fig-conv-B} show that the CG
scheme converges also in this case, albeit significantly slower than in 
Case~A -- about 95 iterations are needed instead of 60. The error maps, 
in the left column of Fig.~\ref{fig-B}, show the W area in stark contrast
to the rest of the sky during the initial iterations. At iteration 20 the 
errors in the W area are still very significant, and have the opposite sign
of their initial values. From iteration 35 and onwards, the errors in the
W area are typically smaller than in the rest of the sky, and in iteration
60 the solution appears to have converged everywhere. However, as 
shown by the truncation error maps in the right column of Fig.~\ref{fig-B}, 
the errors in the W area continue to decrease at least up to iteration 90.
We consider the solution converged from iteration 95, at which point the
parallax updates in the W area are about $10^{-5}~\mu$as. 

That the high contrast in weight between the W area and the rest of the sky
in Case~B has had no negative effect on the solution is more clearly seen in 
Table~\ref{table:2}, which compares the average and RMS parallax errors,
inside and outside of the W area, for the converged solutions in Case~A1 
and Case~B. First of all it can be noted that the average parallax errors 
in all cases are insignificant, i.e., consistent with the given RMS errors
and the assumption that the errors are unbiased.%
\footnote{For example, in Case~B the average error in the W area is expected
to have a standard deviation of $1.88051/\sqrt{504}=0.08376~\mu$as. 
The actual value, $-0.06747~\mu$as, deviates from 0 by only 0.8 standard 
deviations and is therefore not significant.} The slightly negative averages
inside the W area and slightly positive averages in the rest of the sky are
a random effect of the particular noise realization used in these tests, 
and cannot be interpreted as a bias. Secondly, it can be noted that both 
the average parallax error and the RMS parallax error inside the W area 
are reduced roughly in proportion to the observational errors (i.e., by a 
factor $\sim$5), while the errors outside of the W area are very little
affected. This is just as expected in the ideal case that the weight
contrast is correctly handled by the solution. 

\begin{figure}[t!]
\begin{center}
\includegraphics[width=88mm]{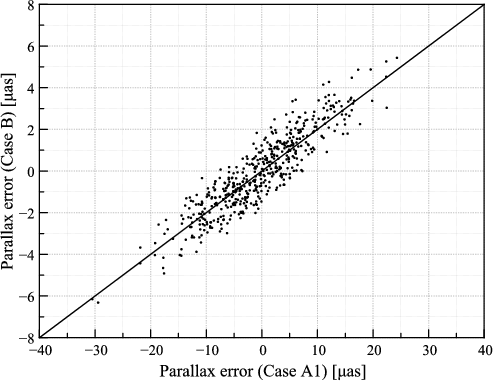}
\caption{Comparison of the individual parallax errors of the 504 sources 
in the W area, from the solution in Case~A1 (horizontal axis) and Case~B 
(vertical axis).}
\label{fig-w}
\end{center}
\end{figure}

\begin{figure*}[t!]
\begin{center}
\centerline{\includegraphics[]{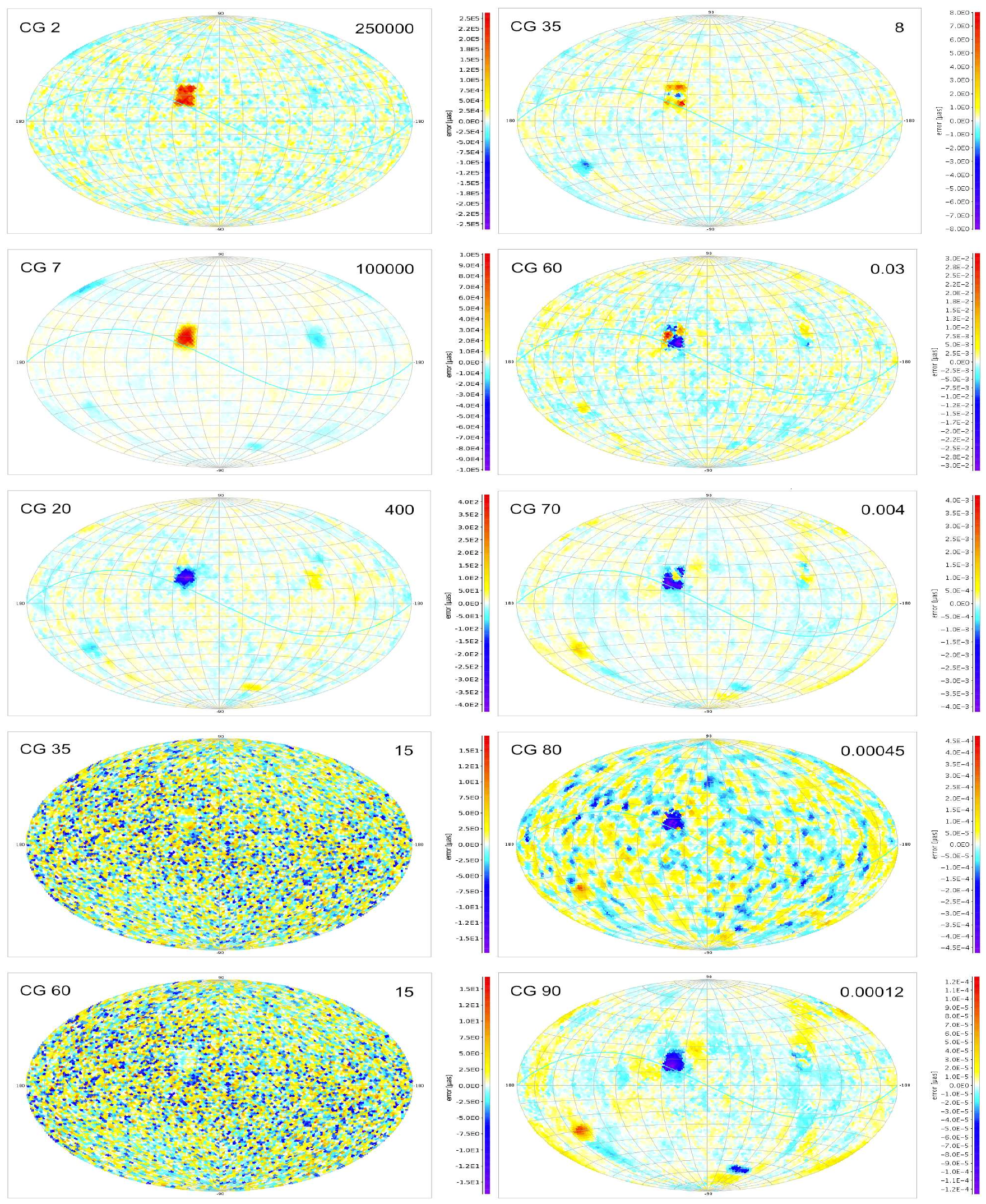}}
\caption{Parallax errors and truncation errors for 
test case B (non-uniform weight distribution). The error maps
in the left column show the differences between the parallax values
in selected iterations and their true values. The truncation error
maps in the right column show the difference between the
parallax values in selected iterations and the converged
values (in iteration 100).   
See Fig.~\ref{fig-A0} for further explanation of the diagram layout. 
}
\label{fig-B}
\end{center}
\end{figure*}

The test cases A1 and B use the same seed for the random observation
errors, which therefore strictly differ by a factor 5 in the W area between
the two cases. This allows a very detailed comparison of the results.
For example, the marginally smaller RMS error outside of the W area in
Case~B ($7.25466~\mu$as) compared to Case~A1 ($7.25570~\mu$as)
is probably real and reflects the improved attitude determination 
in Case~B (thanks to the more accurate observations in the W area),
which benefits also some sources that are not in the W area.
Figure~\ref{fig-w} is a comparison of the individual parallax errors
in the W area
from the two solutions. The diagonal line has a slope of 0.2, equal to
the ratio of the observation errors in the W area between Case~B and 
Case~A1. The diagram suggests that, to a good accuracy, the final 
parallax errors scale linearly with the corresponding observation errors.

\begin{table}
\caption{Average and RMS parallax errors in Case~A1 (where all observations have 
the same standard deviation) and Case~B (where the observations in the W area
have a factor 5 smaller noise). The numbers following the $\pm$ symbol are the
RMS parallax errors. All errors are expressed in $\mu$as.}
\label{table:2}
\centering
\begin{tabular}{lcc}
\hline\hline
\noalign{\smallskip}
Solution & W area & non-W area\\
 & (504 sources) & (99\,496 sources)\\
\noalign{\smallskip}
\hline
\noalign{\smallskip}
Case~A1 (CG\,60) & $-0.45046\pm 8.33135$ & $+0.01341\pm 7.25570$\\
Case~B (CG\,95) & $-0.06747\pm 1.88051$ & $+0.01419\pm 7.25466$\\
\noalign{\smallskip}
\hline
\end{tabular}
\end{table}

\subsection{Definition of a convergence criterion}
\label{sec-conv-crit}

The iteration loops in the SI and CG schemes are set to run for a 
given number of iterations. We now turn to the question how to 
define a convergence criterion, i.e., to determine when to stop the 
iterations. 

In standard implementations of the CG algorithm it is customary to
stop iterating when the norm of the residual vector 
$\vec{r}=\vec{b}-\vec{N}\vec{x}$ is less than some pre-defined
small fraction $\varepsilon$ of the norm of $\vec{b}$
\citep[e.g.,][]{golu+96}. The tolerance $\varepsilon$ must not be
smaller than the unit roundoff error of the floating point arithmetic
used ($2^{-52}\simeq 2\times 10^{-16}$ in our case, using double
precision), but in practice it may have to be many times larger in order
to accommodate the accumulated roundoff errors when 
computing the residual vector. This is especially the case when the
number of unknowns is very large, as in the present application.
The choice of $\varepsilon$ is therefore not trivial: a slightly too
small value would not terminate the iterations, while a slightly too
large value may, as we have seen, result in undesirable truncation 
errors. Ideally we want a convergence criterion that effectively
ensures that we have reached the full accuracy permitted by the 
finite-precision arithmetic.

In this context it is worth pointing out that AGIS is only one step of 
Gaia's data reduction, and that AGIS will be run
many times during the data reduction process. Indeed the output from
AGIS will be used to improve other calibration processes (line spread 
functions, photometry, etc.) which in turn can be used to improve the
astrometric solution. Since 
AGIS is thus part of an outer iteration loop involving several other
calibration processes, it may not be useful to enforce a very strict 
convergence criterion for AGIS until at the very last few outer iterations.
In other words, as long as the other calibrations are not well settled, 
we can live with slightly non-converged astrometric solutions. In
the final outer iteration, the astrometric solution should be driven to
the point where the updates are completely dominated by numerical 
noise. The criteria discussed here have that aim. We consider in the
following only the CG scheme because of its superior convergence
properties. 

In the previous analysis we have studied the convergence in terms 
of the updates $\vec{d}_k$, error vectors $\vec{e}_k$,
and truncation errors $\vec{\epsilon}_k$. The error vectors are of course not 
known for the real mission data, and the truncation errors only
become known after having made many more iterations than strictly 
necessary, and are therefore not useful in practice. The convergence 
criterion could however be based on the updates or various other 
quantities defined in terms of the design equation residuals 
$\vec{s}_k$ or normal equation residuals $\vec{r}_k$ 
(cf.\ Sect.~\ref{sec-iter}).

Based primarily on a visual inspection of the various diagrams, including 
the convergence plots (Figs.~\ref{fig-conv-A0}, \ref{fig-conv-A1}, 
\ref{fig-conv-A2}, \ref{fig-conv-B}) and the parallax truncation errors 
maps (Figs.~\ref{fig-A1a}, \ref{fig-A2}, \ref{fig-B}), it was concluded 
that about 60, 60, 69 and 95 iterations were required for full convergence 
of the CG scheme in the four test cases A0, A1, A2 and B. An ideal 
convergence criterion should tell us to stop at about these points,
and it should also be robust against changes in the number of sources,
their distribution on the celestial sphere, and the weight distribution
of the observations. It is not possible to explore the robustness issue
in this paper, and we therefore concentrate on finding some plausible
candidate criteria.

\begin{figure}[t!]
\begin{center}
\centerline{
\includegraphics[width=88mm]{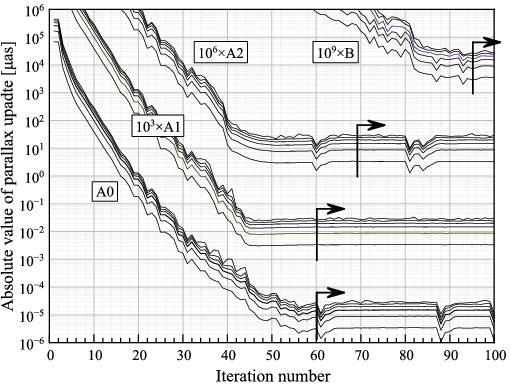}
}
\caption{Statistics of the parallax updates for the CG solutions 
in Case~A0, A1, A2 and B. In each Case the fives curves show, 
from bottom up, the quantiles $q_{0.5}$ (median), $q_{0.9}$, 
$q_{0.99}$, $q_{0.999}$, and $q_{0.9999}$ of the absolute values 
of the parallax updates in each iteration. The thick vertical lines 
with arrows indicate the iterations where the solutions had effectively
converged according to the truncation errors maps. For better
visibility the curves in Case~A1, A2 and B have been shifted upwards
by 3, 6 and 9~dex, respectively.}
\label{fig-updates}
\end{center}
\end{figure}

\subsubsection{Criteria based on the parallax updates}

Among the five astrometric parameters, the parallaxes are especially 
useful for monitoring purposes, because they are not affected by a 
possible frame rotation between successive iterations. It would therefore 
seem natural to define a convergence criterion in terms of some statistic 
of the parallax updates. In the convergence plots we have plotted the 
RMS value of the updates. However,
it is possible that the updates for a small fraction of the sources
(e.g., those with high statistical weights as in the W area of Case~B)
converges less rapidly than the bulk of the sources, and that the
overall standard width of the updates is therefore not the best 
indicator. Instead we will consider quantiles of the absolute values 
of the parallax updates such as $q_{0.999}$ (that is, 99.9\% of the 
absolute updates are less than $q_{0.999}$). 

\begin{figure*}[t!]
\begin{center}
\centerline{
\includegraphics[width=0.98\columnwidth,height=0.46\columnwidth]{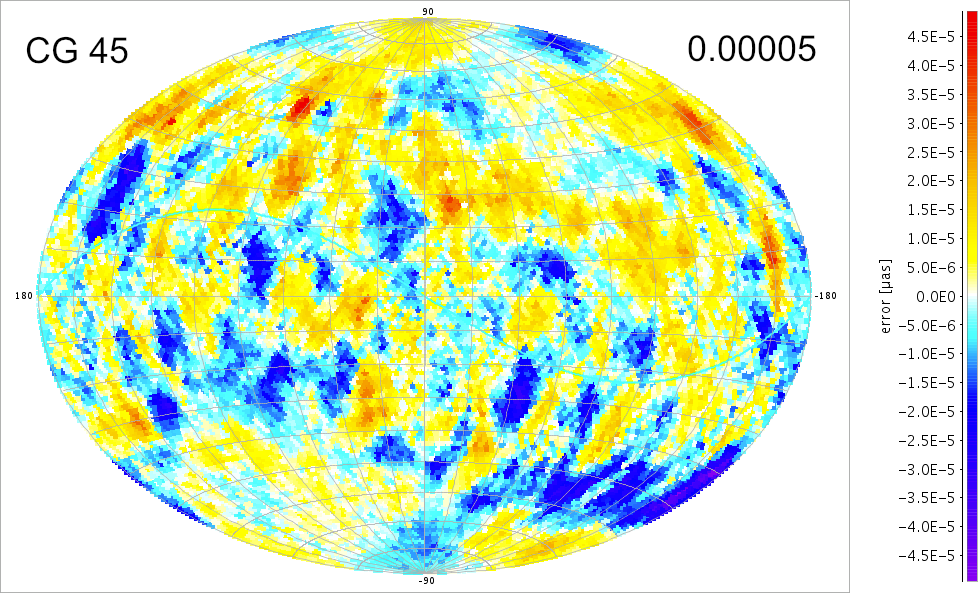}
\hspace{8mm}
\includegraphics[width=0.98\columnwidth,height=0.46\columnwidth]{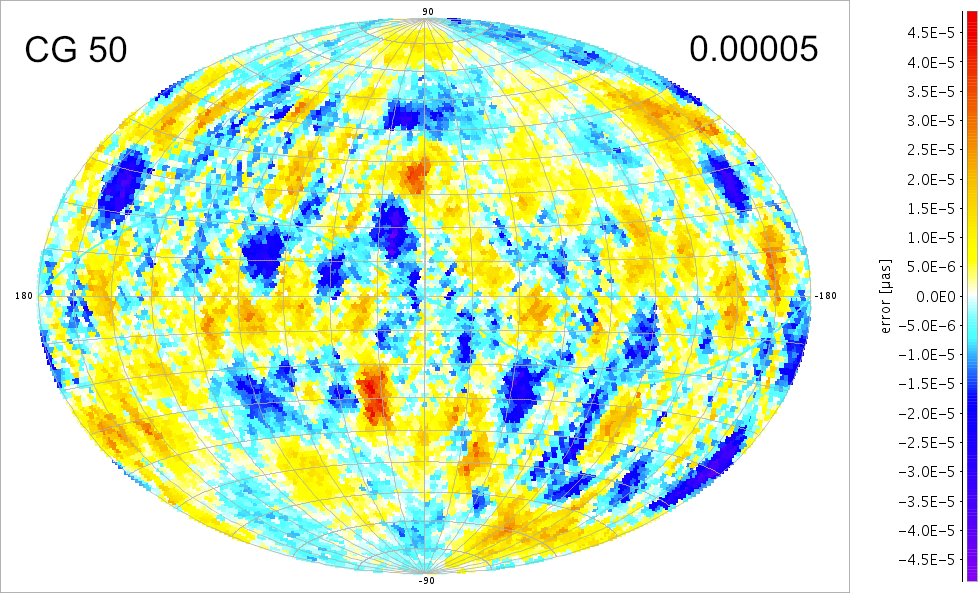}
}
\caption{Truncation error maps for iteration 45 of Case~A1 (left)
and iteration 50 of Case~A2 (right). At these points the parallax
updates have reached their final levels according to Fig.~\ref{fig-updates}, 
but these maps show that the solutions are not quite converged.}
\label{fig-updates-tem}
\end{center}
\end{figure*}

Figure~\ref{fig-updates} summarises the evolution 
of selected quantiles of the absolute parallax updates for the CG 
solutions in the four test cases. The arrows indicate the
first converged iterations according to the previous discussion.
In all cases, the parallax updates eventually reach a final level,
e.g., $\simeq 2\times 10^{-5}~\mu$as for $q_{0.999}$. Remarkably,
however, at least in Case~A1 and A2 this level is reached well before
convergence (e.g., at iteration 45 in Case~A1 and 50 in Case~A2).
At these points, the truncation error maps still contain significant
large-scale features with amplitudes of about $5\times 10^{-5}~\mu$as
(Fig.~\ref{fig-updates-tem}). The same conclusion is reached whatever
quantile is considered. It thus appears that the parallax updates alone 
are not sufficient to define a criterion for the full convergence.
On the other hand, it appears that the magnitude of the updates,
\emph{before} they have reached their final levels, gives a good 
indication of the magnitude of the truncation errors at that point.

\begin{figure*}[t!]
\begin{center}
\centerline{
\includegraphics[width=88mm]{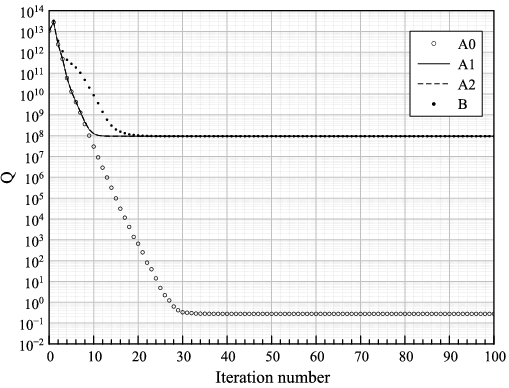}
\hspace{4mm}
\includegraphics[width=88mm]{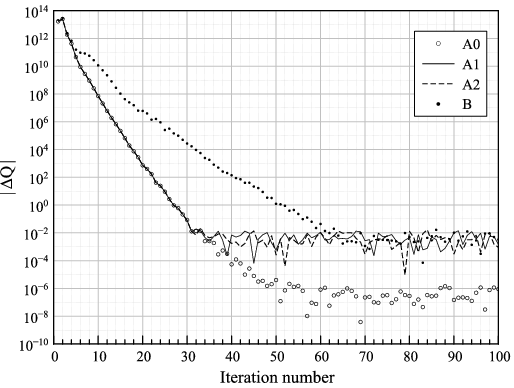}
}
\caption{\emph{Left:} Evolution of $Q_k=\|\vec{s}_k\|^2$ for the CG solutions 
in Case~A0, A1, A2 and B. \emph{Right:} Evolution of the absolute value of
$\Delta Q_k=Q_{k-1}-Q_k$ for the same solutions.}
\label{fig-Q}
\end{center}
\end{figure*}

\begin{figure*}[t!]
\begin{center}
\centerline{
\includegraphics[width=88mm]{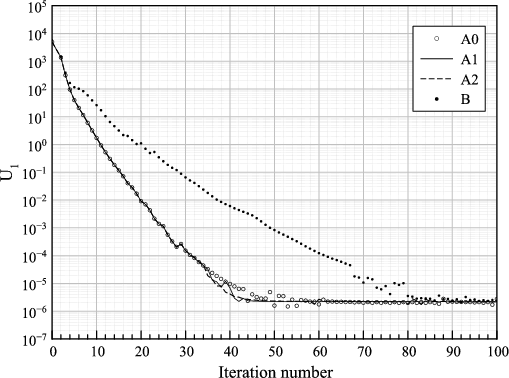}
\hspace{4mm}
\includegraphics[width=88mm]{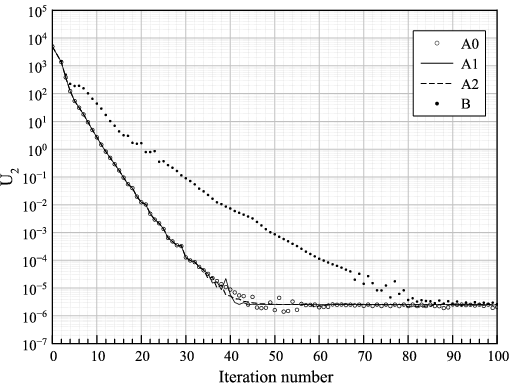}
}
\caption{\emph{Left:} Evolution of $U_1=(\rho_k/n)^{1/2}$ for the CG solutions 
in the four test cases. \emph{Right:} Evolution of 
$U_2=(\alpha_k\rho_k/n)^{1/2}$ for the same solutions.}
\label{fig-U}
\end{center}
\end{figure*}

\begin{figure*}[t!]
\begin{center}
\centerline{
\includegraphics[width=88mm]{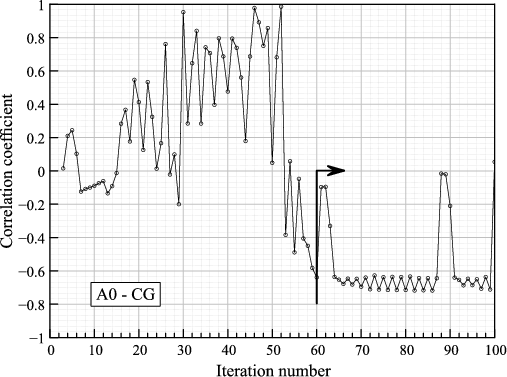}
\hspace{4mm}
\includegraphics[width=88mm]{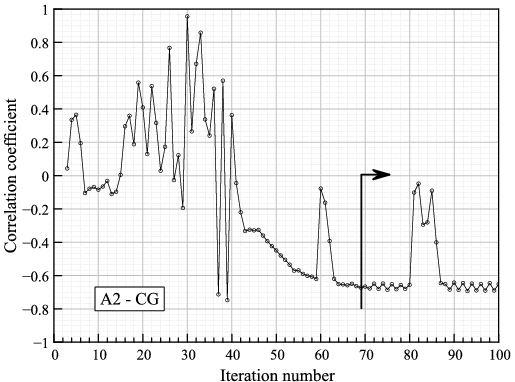}
}\vspace{2mm}
\centerline{
\includegraphics[width=88mm]{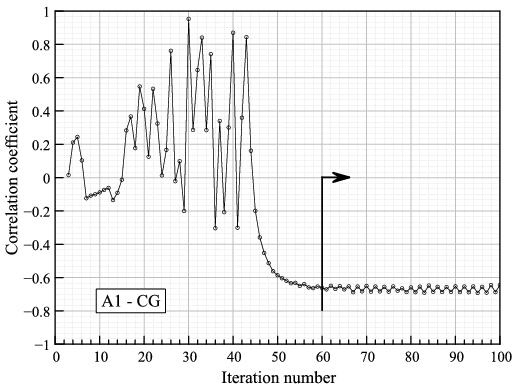}
\hspace{4mm}
\includegraphics[width=88mm]{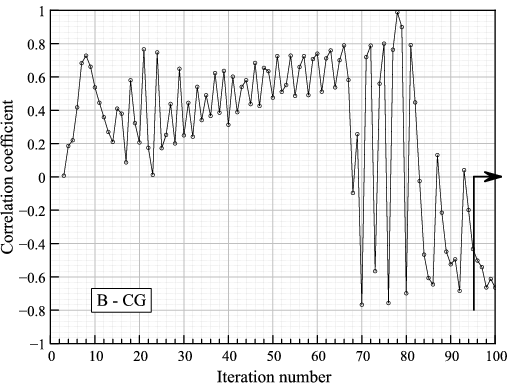}
}
\caption{Correlation coefficient $R_k$ between successive parallax updates
of the CG solutions in Case~A0, A1, A2 and B.
The thick vertical lines with arrows indicate the iterations where 
the solutions had effectively
converged according to the truncation errors maps.}
\label{fig-updatesCorr}
\end{center}
\end{figure*}

\subsubsection{Criteria based on residual vector norms}

Independent of the kernel and iteration schemes, we know for each iteration 
the three vectors $\vec{d}_k$ (updates), $\vec{s}_k$ (design equation 
residuals), and $\vec{r}_k$ (normal equation residuals). There are a number 
of different vector norms that can be computed from these, taking into account
different possible metrics. Ideally, we are looking for a single scalar quantity
that is theoretically known to decrease as long as the solution improves. 

In the CG scheme the square of the norm of the design equation residuals, 
$Q_k=\|\vec{s}_k\|^2$, should be non-increasing according to 
Eq.~(\ref{eq:CGssr}). After convergence, it is expected to reach a value
of the order of $\nu=m-n$, where $m$ is the number of observations and
$n$ the number of unknowns. In our (small-scale) test cases we have
$\nu\sim 10^8$. Figure~\ref{fig-Q} (left) shows that the test cases that
contain observation noise (A1, A2 and B) reach this level in some 15--25
iterations; in the noise-less case (A0) a much lower plateau is reached in 
about 30 iterations. Although not visible in the left diagram, $Q_k$ 
continues to decrease for many more iterations, as shown in the right
diagram of Fig.~\ref{fig-Q}, where the absolute values of 
$\Delta Q_k=Q_k-Q_{k-1}$ are plotted for the same solutions.%
\footnote{$|\Delta Q_k|$ is plotted rather than $\Delta Q_k$ since, due 
to rounding errors, $\Delta Q_k<0$ in many of the later iterations 
(starting at $k=53$, 36, 37 and 69 in Case~A0, A1, A2 and B, respectively).
The negative $\Delta Q_k$ trigger the reinitialisation of the CG algorithm
described in Sect.~\ref{sec-algo-iter}.}
Unfortunately these values seem to reach a stable level even before
the updates. The design equation residuals therefore do not provide a 
useful convergence criterion.

There is no guarantee that the norms of $\vec{d}_k$ and $\vec{r}_k$ 
decrease monotonically, although in the SI scheme they behave 
asymptotically as described in Sect.~\ref{sec-simple} (exponential
decay). The same statements can be made for the scalar product
\begin{equation}\label{def-rho}
\rho_k = \vec{r}_{k}'\vec{d}_k = \vec{d}_{k}'\vec{K}\vec{d}_k 
= \vec{r}_{k}'\vec{K}^{-1}\vec{r}_k\, ,
\end{equation}
which is non-negative for any positive-definite preconditioner 
$\vec{K}$, and has the advantage of being dimensionless.%
\footnote{$\vec{r}$ and $\vec{d}$ are not dimensionless and 
therefore depend on the units used. Indeed, different components of 
these vectors may have different units -- for example $\vec{d}$ 
contains updates both to positions and proper motions, and unless
the unit of time for the proper motions is carefully chosen, the norm 
of this vector makes little sense.} 
Since the quadratic form in Eq.~(\ref{def-rho}) implies a sum over all 
$n$ parameters, we define the RMS-type quantity 
$U_1\equiv (\rho_k/n)^{1/2}$, which is plotted in Fig.~\ref{fig-U}
(left). 

For the CG scheme we note that $\rho_k$ is already calculated as part of 
Algorithm~\ref{algo:CGinCG1}). An even more relevant quantity could be
\begin{equation}\label{def-q}
\vec{d}_{k}'\vec{N}\vec{d}_k 
= \alpha_k^2\vec{p}_{k}'\vec{N}\vec{p}_k = \alpha_k\rho_k \, , 
\end{equation}
which according to Eq.~(\ref{eq:CGssr}) is the amount by which the
sum of squared residuals $Q\equiv\|\vec{s}\|^2$ is expected to decrease 
in the next iteration (mathematically, therefore, 
$\alpha_k\rho_k=\Delta Q_{k+1}$). Since $\vec{N}$ 
is the inverse of the formal covariance of the parameters, 
$U_2\equiv (\alpha_k\rho_k/n)^{1/2}$ has a simple interpretation: 
it is the RMS update defined in units of the statistical errors.

Figure~\ref{fig-U} shows the evolution of $U_1$ and $U_2$ for the CG 
solutions in the four test cases. There is in practice little difference
between $U_1$ and $U_2$, which merely reflects the circumstance that
$\alpha_k$ is of the order of unity throughout the CG iterations.
Moreover, the plots in Fig.~\ref{fig-U} are are quite similar to those 
of the parallax updates in Fig.~\ref{fig-updates}, and therefore
no more useful for defining a convergence criterion.

\subsubsection{Criteria based on the correlation of successive updates}

The conclusion from preceding sections is that various simple statistics 
based on the updates and/or residuals of the \emph{current} iteration are 
insufficient to indicate that the solution has reached full numerical 
accuracy. In particular, none of the above criteria indicate the need to 
continue iterating after iteration 45 in Case~A1, and 50 in Case~A2. 
Yet, inspection of the truncation errors maps in Fig.~\ref{fig-updates-tem}
clearly shows the need for additional iterations. The truncation errors
in the subsequent iterations tend to be similar, only with reduced 
amplitude. This can perhaps be understood as a consequence of
the frequent reinitialisation of the CG algorithm in this regime. In the
limit of constant reinitialisation, Algorithm~\ref{algo:CGinCG1} becomes
equivalent to the SI scheme (Algorithm~\ref{algo:simpleItInCG1}), in 
which the truncation errors tend to decay exponentially. In this
situation the updates also decay exponentially, and therefore have
a strong positive correlation from one iteration to the next. This
suggests that we should look at the correlation between the updates
in \emph{successive} iterations as a possible convergence criterion. 

Figure~\ref{fig-updatesCorr} shows the evolution of the correlation 
coefficient between successive parallax updates $\vec{\delta\varpi}_k$,
\begin{equation}
R_k = \frac{{\vec{\delta\varpi}_k}'{\vec{\delta\varpi}_{k-1}}}%
{({\vec{\delta\varpi}_k}'{\vec{\delta\varpi}_k})^{1/2}
({\vec{\delta\varpi}_{k-1}}'{\vec{\delta\varpi}_{k-1}})^{1/2}}\, ,
\end{equation}
in the four solutions. As in Fig.~\ref{fig-updates}, the arrows indicate
the points where convergence had been reached according to the
discussion above. It is seen that $R_k$ changes from predominantly
positive to predominantly negative values roughly at the points 
when the parallax updates (Fig.~\ref{fig-updates}) or 
$U_1$ and $U_2$ (Fig.~\ref{fig-U}) reach their minimum values
set by the numerical noise. Significantly, however, $R_k$ 
\emph{continues to decrease} beyond these points, 
reaching a roughly constant level $R\simeq -0.67$ at the point 
of convergence.

\begin{figure}[t!]
\begin{center}
\centerline{
\includegraphics[width=88mm]{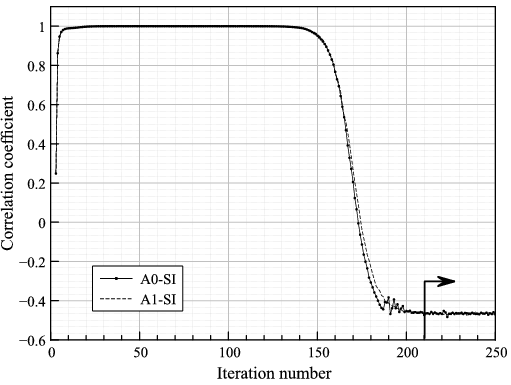}
}
\caption{Correlation coefficient between successive parallax updates
of the SI solutions in Case~A0 and A1. The thick vertical line with 
an arrow indicates where the solutions had effectively
converged according to the truncation errors maps.}
\label{fig-corrSI}
\end{center}
\end{figure}

As already mentioned, the CG scheme more or less reverts to the SI
scheme in the final iterations, due to the frequent reinitialisations.
However, the evolution of the correlation coefficient in
Fig.~\ref{fig-updatesCorr} is partly obscured by the irregularity of 
the reinitialisations -- for example, the sudden rise in $R_k$ at 
iteration 60 and 81 in Case~A2, and at iteration 87 and 93 in Case~B,
seem to be related to the fact that no reinitialisation occurred two
iterations earlier (while otherwise reinitialisation was the rule in this
regime). For comparison we show in Fig.~\ref{fig-corrSI} the evolution 
of $R_k$ in the SI scheme applied to Case~A0 and A1. Here the behaviour
is much more regular, and the correlation coefficient
reaches a stable value of $\simeq -0.47$ from iteration 210,
at which point the solutions had converged according to 
Figs.~\ref{fig-conv-A0} and \ref{fig-conv-A1}.

\subsubsection{Conclusion concerning convergence criteria}

The previous discussion shows how difficult it is to define a reliable 
convergence criterion that is sufficiently strict according to our aims.
Fortunately, as already pointed out, full numerical convergence is
only required in the very final outer processing loop (which includes
many other processes apart from the final astrometric core solution).
For that purpose a combination of the above criteria might be appropriate,
i.e., requiring numerically small updates combined with a correlation 
coefficient that has settled to some negative value. In any case it
will be wise to carry out a few extra iterations after the formal criterion 
has been met.

For the provisional astrometric solutions, where full numerical convergence
is not required, it will be sufficient to stop the CG iterations when the
RMS parallax update or some residual norm such as $U_1$ or $U_2$ 
is below some fixed tolerance (of the order of $0.01~\mu$as and
$10^{-3}$, respectively).

\section{CG implementation status in AGIS\label{sec-agis}}

The AGISLab results presented in this paper, as well as numerous 
other experiments covering a range
of different input scenarios (number of stars, initial noise level
of the unknowns, etc.) convincingly demonstrate the general 
superiority of the CG scheme over SI in terms of convergence rate
and its ability to more quickly remove correlated errors from the solution.
This practical confirmation of the theoretical arguments 
(Sect.~\ref{sec-simple} and \ref{sec-prop-cg}) was an 
important pre-requisite for supporting CG
also in the AGIS framework. This has been done by now,
such that both the SI and CG scheme are available in AGIS
with the same functionality and fidelity as in AGISLab.

The implementation of the core CG scheme is rigorously equivalent
to Algorithm~\ref{algo:CGinCG1} with small but conceptually
irrelevant modifications to better match the existing way of
how the iterations are organized in AGIS. The same is true
for the Gauss--Seidel preconditioner kernel.
The concrete accumulation of design equations is done somewhat
differently to how it is specified in Algorithm~\ref{algo:kernelGSinCG1},
however, the resulting normal equations of the preconditioner $\vec{K}_2$
in Eq.~(\ref{eq:Kform}), viz.\ $\vec{N}_s$ (actually one per source)
and $\vec{N}_a$ are again strictly equivalent to what a faithful
implementation of the algorithm, as in AGISLab, yields.

The algorithms in this paper implicitly assume an underlying all-in-memory
software design which is true for AGISLab but not for AGIS.
Owing to the large data volumes that will have to be processed for
the real Gaia mission (Sect.~\ref{sec-agislab}) AGIS is by necessity a
distributed system \citep{WOM2010}
capable of executing parallel processing threads
on a large number (hundreds to thousands) of independent CPU
cores.
As an example, the accumulation of source and attitude normal equations
is done in different processes running on different CPUs.
Hence, the loop in lines~\ref{alg-gsk-10}--\ref{alg-gsk-12} of
Algorithm~\ref{algo:kernelGSinCG1}, which adds the contribution
of all observations of source $i$ to the attitude normal equations,
cannot be done directly in the same process that treats source $i$.
This complicates matters considerably and, inevitably, leads to different
realizations of CG in AGIS and AGISLab.

Extensive comparisons between AGIS and AGISLab were performed
to ensure the mathematical and numerical equivalence and correctness
of the two
implementations. The tests in AGIS were done using a simulated
data set with 250\,000 stars isotropically distributed across the sky
and very conservative starting conditions for the unknowns
with random and systematic errors of several 10~mas.
A comparable configuration (scaling factor $S$, etc.) was chosen for
AGISLab.

The AGIS results fully confirmed all findings of Sect.~\ref{sec-casea},
notably the important point that CG and SI converge to solutions
which are identical to within the expected numerical limits. Moreover,
a direct comparison of various key parameters as a function of
iteration number, e.g., astrometric source and attitude parameter
errors and updates, solution scalars of CG like $\alpha$, $\beta$, and
$\rho$ (see Algorithm~\ref{algo:CGinCG1}), showed a satisfying agreement
between corresponding AGIS and AGISLab runs. Remaining differences
are at the level of 1 iteration (e.g., the parallax error
reached in AGISLab in iteration $k$ is reached or surpassed in AGIS
not later than in iteration $k+1$ for all values of $k$), and have been
attributed to not
using exactly the same input data (AGISLab uses an internal on-the-fly
simulator). This successfully concluded the validation of the CG implementation
in AGIS.

It is clearly expedient to employ CG as much as possible; however,
in practice with the real mission data we anticipate that a
hybrid scheme consisting of alternating phases of SI and CG
iterations will be needed. The reason is the necessity to identify
and reject outlier observations which is a complex process done through
observation weighting in AGIS. As long as the solution has not converged, 
these weights vary from one iteration
to the next, which means that a slightly different least-squares problem
is solved in every iteration. While this has no negative impact on the very
robust SI method, we have observed that it does not work at all in the case of
CG. This is not surprising and in fact expected as the changing weights lead
to a violation of the conjugacy constraint (see Sect.~\ref{sec-prop-cg})
which is crucial for CG. Hence, we are envisaging a mode in which
AGIS starts with SI iterations up to a point where the weights
have stabilized to a given degree, then activate CG, followed by
perhaps another SI phase to refine the weights further, then again CG,
etc. This will probably make the automatic reinitialization of CG
obsolete. The hybrid SI--CG scheme is
a further development step in AGIS, and a more detailed discussion of
relevant aspects and results is deferred to a future paper.

\section{Conclusion\label{sec-conc}}

We have shown how the conjugate gradient algorithm, with a Gauss--Seidel 
type preconditioner, can be efficiently implemented within the already
existing processing framework for Gaia's Astrometric Global Iterative Solution 
(AGIS). This framework was originally designed to solve the astrometric
least-squares problem for Gaia using the so-called Simple Iteration (SI) 
scheme, which is intuitively straightforward but computationally inefficient. 
The conjugate gradient (CG) scheme, by using the same kernel operations
as SI, takes about the same processing time per iteration but requires a 
factor 3--4 fewer iterations. Both schemes have been extensively
tested using the AGISLab test bed, which allows to perform scaled-down 
and simplified
simulations of Gaia's astrometric observations and the associated least-squares
solution, using (in our case) $10^5$ sources and a total of about 2~million 
source and attitude unknowns.

To within the numerical noise of the double-precision arithmetic,
corresponding to $<10^{-5}~\mu$as in parallax, the SI and CG schemes 
converge to identical solutions. In the case when no observational noise 
was added to the simulated observations, the solutions agree with the
true values of the astrometric parameters to within the numerical noise. 
Thus we conclude that the iterative method provides the correct 
solution to the least-squares problem, provided that a sufficient number
of iterations is used (full numerical convergence reached). As theoretically 
expected, the resulting solution is completely insensitive to the initial 
values of the astrometric parameters, although the rate of convergence
may depend on the initial errors.

Although the SI and CG schemes eventually reach the same solution,
to within the numerical precision of the computations, the truncation errors
obtained by prematurely stopping the iteration have quite different 
character in the two schemes. In the SI scheme the truncation error maps 
are often strongly correlated over large parts of the celestial sphere, while 
in the CG scheme there is less spatial correlation. Thus, the CG scheme
not only converges faster than SI, but the truncation errors at a given level
of the updates are considerably more `benign' in terms of large-scale 
systematics.

Most of the numerical experiments use a highly idealised, uniform 
distribution of sources, and a uniform level of the standard error of 
the observations. However, it has been demonstrated that the solution
works flawlessly also in the case when the observations have a much
larger weight in a small part of the sky (representing, for example, a 
bright stellar cluster). Although such a situation needs more iterations,
the converged solution correctly reflects the weight distribution of
the observations -- i.e., the accuracy of the astrometric parameters
of the cluster stars is increased roughly in proportion to their increased
observational accuracy, without any noticeable negative impact on the 
results for other stars.

We have stressed the need to drive the iterations to full numerical
convergence, at least in the final astrometric solution for the Gaia 
catalogue. This is important in order to avoid that the final catalogue
contains truncation errors that are unrelated to the mission and
satellite itself, but merely caused by inadequate processing. Such
truncation errors could mimic `systematic errors' in the catalogue.
The Gaia catalogue will certainly not be free of systematic errors, but
at least we should insist that they are not produced by prematurely 
stopping the astrometric iterations. However, achieving full numerical
convergence may require many iterations beyond the point where
simple metrics indicate convergence. It is in fact quite difficult to 
define a numerically strict convergence criterion, although we have
found that the correlation between updates in successive iterations 
may provide a useful clue. At the point where the updates have not
yet settled at their final level, the magnitude of the updates gives a
good indication of the remaining truncation errors. If one does not
insist on full numerical convergence, it is therefore relatively safe
to stop iterating when the updates have reached a sufficiently low
level, say below 0.01~$\mu$as. 

The CG algorithm described in this paper only considers the two major
processing blocks in AGIS, namely the determination of the source and
attitude parameters. This restriction was intentional, in order to
simplify the description and numerical testing. At the same time, 
the successful disentangling of the source and attitude parameters 
is the key to a successful solution, as shown by numerous experiments.
Nevertheless, the proposed algorithm is readily extended to the fully 
realistic problem that includes calibration and global parameters, and has 
in fact been realised in this form and successfully demonstrated in the 
current AGIS implementation at the Gaia data processing centre in ESAC 
(Madrid).

Our aim has been to investigate the applicability of the CG algorithm 
for solving Gaia's astrometric least-squares problem efficiently
within the AGIS framework. To this end we have 
considered a scaled-down and highly idealized version of the problem 
where many detailed complications in the real Gaia data are ignored. 
Nevertheless, within the given assumptions, we have successfully
demonstrated how the CG algorithm can be adapted to the astrometric core 
solution. Moreover, by means of numerical simulations we have shown
that the numerical accuracy achieved with this method is high enough 
that it will not be a limiting factor for the quality of the astrometric 
results.

\begin{acknowledgements}
This work was carried out in the context of the European Marie-Curie research
training network ELSA (contract MRTN-CT-2006-033481), with additional support
from the Swedish National Space Board and the European Space Agency.
We thank the referee, Dr.~F.~van Leeuwen, for several constructive comments 
which helped to improve the original version of the manuscript.
The algorithms were typeset using the \LaTeX\ {\sl algorithms} package.
\end{acknowledgements}

\bibliographystyle{aa}
\bibliography{aa17904-11}

\appendix

\section{Efficiency of the least-squares method}
\label{sec-gauss-markoff}

This appendix reviews some important properties of the least-squares
method, which motivate its present application.

The estimation problem associated with the astrometric core solution has 
the following characteristics: (i) within the expected size of the errors, it is
completely linear in terms of the adjusted parameters or unknowns; (ii) the
observational errors are unbiased, (iii) uncorrelated, and (iv) of known 
standard deviation. Property (i) follows from the small absolute sizes of the
errors; (ii) assumes accurate modelling at all stages of the data analysis; (iii)
follows from the Poissonian nature of the photon noise being by far the
dominating noise source; and (iv) is ensured by the estimation method
applied to the individual photon counts in the raw data. Dividing each
observation equation by the known standard deviation of the observation
error thus results in the standard linear model 
$\vec{h}=\vec{M}\vec{x}+\vec{\nu}$, where the vector of random
observation noise $\vec{\nu}$ has expectation 
$\text{E}(\vec{\nu})=\vec{0}$
and covariance $\text{E}(\vec{\nu\nu'})=\vec{I}$. This is 
equivalent to the overdetermined set of design equations introduced 
in Sect.~\ref{sec-iter}.

The Gauss--Markoff theorem \citep[e.g.,][]{book:bjork-1996} states that if
$\vec{M}$ has full rank, then the best linear unbiased estimator (BLUE) 
of $\vec{x}$ is obtained by minimising the sum of squares 
$Q\equiv\|\vec{h}-\vec{M}\vec{x}\|^2$. This is known as the ordinary
least-squares estimator $\hat{\vec{x}}$ and can be computed by solving 
the normal equations $\vec{M}'\vec{M}\hat{\vec{x}}=\vec{M}'\vec{h}$.
For any of the unknowns $x_i$, the theorem implies that the value 
$\hat{x}_i$ obtained by the least-squares method is unbiased 
($\text{E}(\hat{x}_i)=x_i^\text{(true)}$) and that, among all possible 
unbiased estimates that are linear combinations of the data ($\vec{h}$), 
it has the smallest variance.%
\footnote{When $\vec{M}$ is rank deficient, as in the present problem
(Sect.~\ref{sec-back}), the theorem still holds for the part of the
solution vector that is orthogonal to the null space.} 
In terms of the estimation errors $\hat{\vec{e}}$ we have 
$\text{E}(\hat{\vec{e}})=\vec{0}$. The formal uncertainties and
correlations of the estimated parameters are given by  the covariance matrix 
$\text{E}(\hat{\vec{e}}\hat{\vec{e}}')=(\vec{M}'\vec{M})^{-1}$.
In practice it is not feasible to calculate elements of this matrix 
rigorously, so approximate methods must be used (Holl et al., in prep.). 

Systematic errors are by definition discrepancies in the estimated values,
i.e., $\text{E}(\hat{\vec{e}})\neq\vec{0}$, because of a lack of details 
in the modelling and/or because the observation noise is not as expected.
We should not confuse these kinds of errors with errors due to the solver.
Indeed, using an iterative solver can lead to truncation errors because the
solver has not been iterated enough, or worse: because it does not converge
toward the solution of the least-squares problem. It is therefore important
to verify that the different iteration schemes, given enough iterations, do
indeed converge to identical solutions.

\section{Conjugate gradient from mathematics to an algorithm}
\label{sec-CG-van}

Although different implementations of the CG algorithm are mathematically 
equivalent, their behaviours may be completely different in a finite-precision 
environment. Algorithm~\ref{algo:CGinCG1} described in Sect.~\ref{sec-algo-iter} 
is based on a scheme given by \citet{vanVorst:03}, which is here reproduced
as Algorithm~\ref{algo:CGvan}. In this appendix we discuss the changes 
introduced to this scheme and their motivation in terms of its  
implementation in the AGIS framework \citep{LL:LL-077}. For brevity we hereafter 
refer to Algorithm~\ref{algo:CGinCG1} as the CG scheme, and to 
Algorithm~\ref{algo:CGvan} as the vdV scheme.

\begin{algorithm}[t]
\caption{The conjugate gradient method with preconditioner, as given 
in Fig.~5.2 of \citet{vanVorst:03}, but using our notations.}
\label{algo:CGvan}
\begin{algorithmic}[1]
\STATE{initial guess $\vec{x}_0$}
\STATE{$\vec{r}_0 \leftarrow \vec{b} - \vec{N}\vec{x}_0$}\label{alg7-2}
\FOR{ $k=1,\,2,\,\hdots$ }
	\STATE{$\vec{w}_{k-1} \leftarrow K^{-1}\vec{r}_{k-1}$}\label{alg7-4}
	\STATE{$\rho_{k-1} \leftarrow {\vec{r}_{k-1}}'\vec{w}_{k-1}$}
	\IF{$k=1$}
		\STATE{$\vec{p}_k\leftarrow \vec{w}_{k-1}$}
	\ELSE
		\STATE{$\beta_{k-1} \leftarrow \rho_{k-1}/\rho_{k-2}$}
		\STATE{$\vec{p}_k \leftarrow \vec{w}_{k-1} + \beta_{k-1} \vec{p}_{k-1}$}
	\ENDIF
	\STATE{$\vec{q}_k \leftarrow \vec{N}\vec{p}_k $}\label{alg7-12}
	\STATE{$\alpha_k \leftarrow \rho_{k-1} / ({\vec{p}_k}'\vec{q}_{k})$}
	\STATE{$\vec{x}_k \leftarrow \vec{x}_{k-1} + \alpha_k \vec{p}_k$}\label{alg7-14}
	\STATE{$\vec{r}_k \leftarrow \vec{r}_{k-1} - \alpha_k \vec{q}_k$}
\ENDFOR
\end{algorithmic}
\end{algorithm}

Comparing the two schemes, we note several important differences. First of all, 
the kernel in CG combines the computation of the normal equation residuals
$\vec{r}$ and the solution $\vec{w}$ of the preconditioner equations in 
lines~\ref{alg7-2} and \ref{alg7-4}, respectively, of the vdV scheme. This is
expedient since both computations are based on the setting up and (partially)
solving the same normal equations, as explained in Sect.~\ref{sec-algo-kernel}.
Indeed, each of them requires a loop through all the observations, and doing them
in parallel obviously saves both input/output operations and calculations.

The next important difference is found in line~\ref{alg7-12} of the vdV scheme,
where the vector $\vec{q}$ is introduced by another calculation involving the
normal matrix $\vec{N}$. Taken at face value, this step seems to require another 
loop through the observations in order to compute the right-hand side of the
normal equations for point $\vec{p}$ in solution space. In CG this step is avoided 
by the following device. From line~\ref{alg7-14} in vdV we note that the next update 
of $\vec{x}$ is a scalar $\alpha$ times $\vec{p}$. Now let us tentatively assume 
$\alpha=1$ and compute the new, tentative normal equation residuals 
$\tilde{\vec{r}}$ -- this is done in lines~\ref{alg5-6}--\ref{alg5-7} of the CG 
scheme. At that point we have the residual vector $\vec{r}$ referring to the 
original point $\vec{x}$, and $\tilde{\vec{r}}$ referring to $\vec{x}+\vec{p}$. 
Thus, $\vec{r}=\vec{b}-\vec{N}\vec{x}$ and 
$\tilde{\vec{r}}=\vec{b}-\vec{N}(\vec{x}+\vec{p})=\vec{r}-\vec{N}\vec{p}$
from which we find $\vec{q}=\vec{r}-\tilde{\vec{r}}$. This explains
line~\ref{alg5-8} in CG. Once $\alpha$ has been calculated, the
tentative update in line~\ref{alg5-6} can be corrected in line~\ref{alg5-9}.
Lines~\ref{alg5-10}--\ref{alg5-12} make the corresponding corrections
to $Q$, $\vec{r}$, and $\vec{w}$, so that these quantities hereafter are 
exactly as if the kernel had been computed for the point 
$\vec{x}+\alpha\vec{p}$.

From a purely mathematical point of view these modifications do not
change anything, but for AGIS the trick is essential in order to save
computations, since most of the time is spent setting the 
preconditioner equations at a given point in the solution space.
Calculating $\vec{q}$ as the difference $\vec{r}-\tilde{\vec{r}}$ may 
be numerically less accurate than $\vec{N}\vec{p}$, and this could trigger 
an earlier reinitialization of the CG, but this is a small price to pay for 
the improved efficiency.

\end{document}